\documentclass[journal,twocolumn,10pt]{IEEEtran}
\usepackage{graphicx}
\usepackage{amssymb}
\usepackage{mathtools}
\usepackage{dsfont}
\usepackage{cite}
\usepackage{stfloats}
\usepackage{subfigure}
\usepackage{psfrag}
\usepackage[mathscr]{euscript}
\usepackage{acronym}  % make an acronym
\usepackage{amsmath}
\usepackage{algorithm}
\usepackage[noend]{algpseudocode}
\usepackage{booktabs}
\usepackage{bbm}

\makeatletter
\def\BState{\State\hskip-\ALG@thistlm}
\makeatother

\usepackage{float}
\usepackage{balance}

\acrodef{CCDF}{complementary cumulative distribution function}
\acrodef{CF}{characteristic function}
\acrodef{PPP}{Poisson point processe}
\acrodef{RV}{random variable}
\acrodef{i.i.d.}{independent and identically distributed}
\acrodef{PDF}{probability distribution function}
\acrodef{CDF}{cumulative distribution function}
\acrodef{ch.f.}{characteristic function}
\acrodef{AWGN}{additive white Gaussian noise}
\acrodef{SNR}{signal-to-noise ratio}
\acrodef{LRT}{likelihood ratio test}
\acrodef{DRT}{distance ratio test}
\acrodef{GLRT}{generalized likelihood ratio test}
\acrodef{CRLB}{Cram\'{e}r-Rao lower bound}
\acrodef{CRB}{Cram\'{e}r-Rao bound}
\acrodef{ZZLB}{Ziv-Zakai lower bound}
\acrodef{ZZB}{Ziv-Zakai bound}
\acrodef{LOS}{line-of-sight}
\acrodef{ToF}{time-of-flight}
\acrodef{NLOS}{non-line-of-sight}
\acrodef{GDOP}{geometric dilution of precision}
\acrodef{GPS}{Global Positioning System}
\acrodef{FIM}{Fisher information matrix}
\acrodef{PEB}{position error bound}
\acrodef{SPEB}{squared position error bound}
\acrodef{TOA}{time-of-arrival}
\acrodef{TOF}{time-of-flight}
\acrodef{WSN}{wireless sensor network}
\acrodef{MAC}{medium access control}
\acrodef{RSS}{received signal strength}
\acrodef{WAF}{wall attenuation factor}
\acrodef{TDOA}{time difference-of-arrival}
\acrodef{RF}{radiofrequency}
\acrodef{RTT}{round-trip time}
\acrodef{AOA}{angle-of-arrival}
\acrodef{MF}{matched filter}
\acrodef{ED}{energy detector}
\acrodef{ML}{maximum likelihood}
\acrodef{MSE}{mean-square error}
\acrodef{RMSE}{root-mean-square error}
\acrodef{LEO}{localization error outage}
\acrodef{ppm}{part-per-million}
\acrodef{ACK}{acknowledge}
\acrodef{UWB}{Ultrawide bandwidth}
\acrodef{TNR}{threshold-to-noise ratio}
\acrodef{LS}{least squares}
\acrodef{IR-UWB}{impulse radio UWB}
\acrodef{FCC}{Federal Communications Commission}
\acrodef{TH}{time-hopping}
\acrodef{PPM}{pulse position modulation}
\acrodef{MUI}{multi-user interference}
\acrodef{PDP}{power delay profile}
\acrodef{BPZF}{band-pass zonal filter}
\acrodef{SIR}{signal-to-interference ratio}
\acrodef{SINR}{signal-to-interference-plus-noise ratio}
\acrodef{RFID}{radio frequency identification}
\acrodef{WPAN}{wireless personal area network}
\acrodef{WWB}{Weiss-Weinstein bound}
\acrodef{DP}{direct path}
\acrodef{MF}{matched filter}
\acrodef{MMSE}{minimum-mean-square-error}
\acrodef{SBS}{serial backward search}
\acrodef{SBSMC}{serial backward search for multiple clusters}
\acrodef{NBI}{narrowband interference}
\acrodef{WBI}{wideband interference}
\acrodef{INR}{interference-to-noise ratio}
\acrodef{CR}{channel response}
\acrodef{CIR}{channel impulse response}
\acrodef{CR}{channel  response}
\acrodef{RADAR}{radar}
\acrodef{MUR}{Multistatic radar}
\acrodef{JBSF}{jump back and search forward}
\acrodef{HDSA}{high-definition situation-aware}
\acrodef{RRC}{root raised cosine}
\acrodef{ST}{simple thresholding}
\acrodef{BTB}{Bellini-Tartara bound}
\acrodef{P-Max}{$P$-Max}  %suggestion, use with \acl{P-Max}
\acrodef{MIMO}{multiple-input multiple-output}
\acrodef{MAP}{maximum a posteriori}
\acrodef{FG}{factor graph}
\acrodef{OP}{outage probability}
\acrodef{WED}{wall extra delay}
\acrodef{RMS}{root mean square}
\acrodef{SPAWN}{sum-product algorithm over a wireless network}
\acrodef{MDD}{minimum distance distribution}
\acrodef{MAP}{maximum a posteriori probability}
\acrodef{SAP}{small cell access point}
\acrodef{UE}{user equipment}
\acrodef{MBS}{macro cell base station}
\acrodef{UER}{\ac{UE} Relay}
\acrodef{D2D}{device-to-device}
\acrodef{MBS}{macro base station}
\acrodef{CSI}{channel state information}
\acrodef{OGR}{outage guard region}
\acrodef{FUR}{feasible UER region}
\acrodef{EHR}{energy harvesting region}
\acrodef{EH}{energy harvesting}
\acrodef{D2D-EHSN}{D2D communication provided \ac{EH} small cell network}
\acrodef{D2D-EHHN}{D2D communication provided \ac{EH} heterogeneous network}
\acrodef{3GPP}{3rd Generation Partnership Project}
\acrodef{BS}{base station}
\acrodef{DF}{decode and forward}
\acrodef{CCDF}{complementary cumulative distribution function}
\acrodef{ZF}{zero forcing}
\acrodef{RZF}{regularized zero forcing}
\acrodef{WLLN}{weak law of large number}
\acrodef{SLLN}{strong law of large numbers}
\acrodef{TDD}{Time-division duplex}
\acrodef{EE}{energy efficiency} 
\acrodef{HetNet}{heterogeneous network} 
\acrodef{SCP}{Single Cell Processing}
\acrodef{CBF}{Coordinated Beamforming}
\usepackage{color}
\usepackage{dsfont}
\usepackage{bbm}

\DeclareMathAlphabet{\mathsf}{OML}{cmbr}{m}{it}

\newtheorem{theorem}{\bf Theorem}
\newtheorem{lemma}{\bf Lemma}
\newtheorem{corollary}{\bf Corollary}

\newtheorem{remark}{\bf Remark}

\newtheorem{assumption}{\bf Assumption}

\newcommand{\bd}{\begin{description}}
\newcommand{\ed}{\end{description}}
\newcommand{\be}{\begin{enumerate}}
\newcommand{\ee}{\end{enumerate}}
\newcommand{\bi}{\begin{itemize}}
\newcommand{\ei}{\end{itemize}}
\newcommand{\bl}{\begin{list}}
\newcommand{\el}{\end{list}}
\newcommand{\bt}{\begin{tabbing}}
\newcommand{\et}{\end{tabbing}}

\newcommand{\paperTitle}{ Optimizing Information Freshness in Wireless Networks: A Stochastic Geometry Approach }

\begin{document}

{
\title{\paperTitle}

\author{

	    Howard~H.~Yang, \textit{Member, IEEE}, Ahmed~Arafa, \textit{Member, IEEE}, \\
	    Tony~Q.~S.~Quek, \textit{Fellow, IEEE}, and H. Vincent Poor \textit{Fellow, IEEE}\\
     %  \textit{Singapore University of Technology and Design, Singapore}

% \thanks{Manuscript received Feb. 03, 2018, revised Mar. 31, and May 28, 2018, and accepted May 29, 2018. The associate editor coordinating the review of this letter and
%     approving it for publication was Dr. Chun Tung Chou.

%     This work was supported in part by the MOE ARF Tier 2 under Grant MOE2015-T2-2-104 and in part by the SUTD-ZJU Research Collaboration under Grant SUTD-ZJU/RES/01/2016.}

\thanks{H.~H.~Yang and T.~Q.~S.~Quek are with the Information Systems Technology and Design Pillar, Singapore University of Technology and Design, Singapore (e-mail: howard\_yang@sutd.edu.sg, tonyquek@sutd.edu.sg).

A.~Arafa is with the Department of Electrical and Computer Engineering, University of North Carolina at Charlotte, NC 28223, USA (email: aarafa@uncc.edu).

H.~V.~Poor is with the Department of Electrical Engineering, Princeton University, Princeton, NJ 08544 USA (e-mail: poor@princeton.edu).
}
% \thanks{G.~Geraci is with the Department of Small Cells Research, Nokia Bell Labs, Dublin, Republic of Ireland (e-mail: dr.giovanni.geraci@gmail.com).}
% \thanks{Y.~Zhong is with the School of Electronic Information and Communications, Huazhong University of Science and Technology, Wuhan, P.R. China (email: yzhong@hust.edu.cn).}
% \thanks{H.~H.~Yang and T.~Q.~S.~Quek are with the Singapore University of Technology and Design (e-mail: howard\_yang@sutd.edu.sg, tonyquek@sutd.edu.sg). Y.~Wang is with Nanjing University of Post and Telecommunications (e-mail: wangy1585@163.com). }
}
\maketitle
\acresetall
\thispagestyle{empty}
\begin{abstract}
Optimization of information freshness in wireless networks has usually been performed based on queueing analysis that captures only the temporal traffic dynamics associated with the transmitters and receivers. However, the effect of interference, which is mainly dominated by the interferers' geographic locations, is not well understood.
In this paper, we leverage a spatiotemporal model, which allows one to characterize the age of information (AoI) from a joint queueing-geometry perspective, for the design of
a decentralized scheduling policy that exploits local observation to make transmission decisions that minimize the AoI.
To quantify the performance, we also derive accurate and tractable expressions for the peak AoI. Numerical results reveal that:
$i$) the packet arrival rate directly affects the service process due to queueing interactions,
$ii$) the proposed scheme can adapt to traffic variations and largely reduce the peak AoI,
and $iii$) the proposed scheme scales well as the network grows in size. This is done by adaptively adjusting the radio access probability at each transmitter to the change of the ambient environment.
\end{abstract}

% Note that keywords are not normally used for peerreview papers.
}

\acresetall
%%%%%%%%%%%%%%%%%%%%%%%%%%%%%%%%%%%%%%%%%%%%%%%%%%%%
\section{Introduction}\label{sec:intro}
Fast-growing wireless services like factory automation and vehicular communication, as well as the likes of mobile applications, have imposed a more stringent requirement for the timely delivery of information.
To give an adequate response, network operators need not only understand how the network activities affect the timeliness of information delivery, but more importantly, they need to assert substantial control to enhance transmission.
Recognizing the limitation in conventional performance indicators, e.g., delay or throughput, as not being able to account the ``information lag'' caused by queueing aspects, there emerges a new metric, referred to as the \textit{age of information (AoI)}, which explicitly measures the time elapsed since the last recorded  update was generated.
The notion was originally conceived to maintain timely status update in a standard first-come-first-served (FCFS) queue \cite{KauYatGru:12}.
Soon after that, a host of research has been conducted to investigate different schemes aimed at minimizing the AoI, whereas the results range from controlling the update generating policy \cite{HuaMod:15,Yat:15ISIT,BacSunUys:18,WuYanWu:18,AraYanUlu:18,AraUlu:19,SunUysYat:17}, deploying last-come-first-served (LCFS) queue \cite{CheKua:16ISIT}, to proactively discarding stale packets at the source node \cite{CosCodEph:16}.
Although these works have extensively explored the minimization of the information age on a single-node basis, many fundamental questions, especially those pertaining to large scale networks are not understood satisfactorily. To that end, it spurred a series of studies seeking different approaches, mainly in the form of scheduling protocols, to optimize information freshness in the context of wireless networks \cite{HeYuaEph:16,KadUysSin:16,SunUysKom:18,TalKarMod:18,talak2018optimizing,KadSin:18,HsuModDua:17,BedSunShr:17}.
The problem of finding optimal scheduling protocol, despite being NP hard \cite{HeYuaEph:16}, is shown to possess a solution in terms of a greedy algorithm, which  schedules the link with the highest age to transmit, in a symmetric network \cite{KadUysSin:16}, and the optimality of such a maximum age first policy is shown in \cite{SunUysKom:18}, which provided a general and insightful sample-path proof.
Moreover, depending on whether the channel state is perfectly available \cite{TalKarMod:18} or not \cite{talak2018optimizing}, advanced virtual queue and age-based protocols are proposed.
Further, the scheduling decision can even be made online using the approximation from Markov decision process \cite{HsuModDua:17}, which largely boosts the implementational efficiency.
%However, these models over simplify the wireless channel and lack the ability to track the interference, which differs according to distance between simultaneous transmitters as well as channel gains, thus do not capture the information-theoretic interactions precisely.
However, these models simplify the packet departure process by adopting a Poisson process and do not account for the interference that differs according to the distance between simultaneous transmitters as well as channel gains.
As a result, the space-time interactions are yet to be precisely captured.

By nature, the wireless channel is a broadcast medium. Thus, transmitters sharing a common spectrum in space will interact with each other through the interference they cause.
To understand the performance of communication links in such networks, stochastic geometry has been introduced as a tool by which one can model node locations as spatial point processes and obtain closed form expressions for various network statistics, e.g., the distribution of interference, the successful transmission probability, and the coverage probability \cite{HaeAndBac:09}.
The power of stochastic geometry has made it a disruptive tool for performance evaluation among various wireless applications, including ad-hoc and cellular networks \cite{AndBacGan:11}, D2D communications \cite{YanLeeQue:16}, MIMO \cite{YanGerQue:16}, and mmWave systems \cite{BaiHea:15}.
While such model has been conventionally relying on the \textit{full buffer} assumption, i.e., every link always has a packet to transmit, a line of recent works managed to bring in queueing theory and relax this constraint \cite{GhaElsBad:17,ZhoQueGe:16,YanQue:19,YanWanQue:18}.
The application territory of stochastic geometry is then stretched, allowing one to give a complete treatment for the behavior of wireless links in a network with spatial and temporal dynamics. As a result, the model is further employed to design scheduling policies \cite{ZhoQueGe:16,YanWanQue:18}, study the scaling property in IoT networks \cite{GhaElsBad:17}, and analyze the delay performance in cellular networks \cite{YanQue:19}.

In light of its effectiveness, we leverage a spatiotemporal model as in \cite{ChiElSCon:17,ChiElsCon:19} for the design of a transmission protocol that optimizes information freshness in wireless networks. Particularly, our goal is to develop a policy that decides whether the current transmission attempt from any transmitter shall be approved or not so as to control the cross-network interference level and hence assure the timely delivery of information.
By noticing that centralized protocols, e.g., the ones proposed in \cite{TalKarMod:18,talak2018optimizing}, can incur large communication overhead and do not scale with the network size, we propose a distributed scheduling policy that exploits only local information to control the medium access probability at each node.
Notably, while a few recent attempts to the design and analysis of a scheduling scheme under similar model have been made in \cite{ChiElSCon:17,ZhoQueGe:16,ChiElsCon:19}, the current paper differs from and generalizes these works in two key aspects:
\begin{itemize}
	\item[1)] \textit{Design}: Different from \cite{ChiElSCon:17,ChiElsCon:19,ZhoQueGe:16}, where the channel access probability is universally designed as a single parameter, our approach gives a channel access probability which is a function of local topology and varies among different transmitters.
	\item[2)] \textit{Analysis}: While \cite{ChiElSCon:17,ChiElsCon:19,ZhoQueGe:16} carry out their analysis based on constant channel access probabilities, our analysis characterizes the dynamics of a scheduling policy that changes according to node locations. We also provide an analysis for the peak AoI with dynamic channel access probabilities at different transmitters.
\end{itemize}

\subsection{Approach and Summary of Results}
In this paper, we model the deployment of transmitters and receivers as independent Poisson point processes (PPPs).
The temporal dynamic of AoI is modeled as a discrete-time queueing system, in which we consider the arrival of packets at each transmitter to be independent Bernoulli processes.
Each transmitter maintains an infinite capacity buffer to store the incoming packets, and initiates a transmission attempt at each time slot if the buffer is not empty.
Transmissions are successful only if the signal-to-interference-plus-noise ratio (SINR) exceeds a predefined threshold, upon which the packet can be removed from the buffer.
In order to control the cross-network interference level and hence assure the timely delivery of information, a scheduling protocol is employed at each node to decide whether the current transmission attempt should be approved or not.
Design of the scheduling policy exploits local observation, which is encapsulated via the concept of stopping sets \cite{BacBlaSin:14}, to make the transmission decision, aimed at optimizing the network-wide information freshness.
In order to characterize the performance of our proposed scheme, we derive accurate and tractable expressions for the scheduled channel access probability, the transmission success probability, and the peak AoI.
The analytical results enable us to explore the effect from various network parameters on the AoI and hence devise useful insights for the protocol design.
Our main contributions are summarized below.
\begin{itemize}
\item We propose a decentralized scheduling policy to minimize the information age in a wireless network. The proposed scheme is efficient in the sense that it requires only local information and has very low implementation complexity.

\item We develop an analytical framework that captures the interplay between the geographic locations of information source nodes and their temporal traffic dynamics. Using the framework, we derive tractable expressions for various network statistics by taking into account all the key features of a wireless network, including packet arrival rate, small scale fading and large scale path-loss, random network topology, and spatially queueing interactions.

\item Numerical results show that although the packet arrival rate directly affects the service process via queueing interaction, our proposed scheme can adapt to the traffic variations and largely reduce the peak AoI. Moreover, the proposed scheme can also adequately adjust according to the change of the ambient environment and thus scales well as the network grows in size.
\end{itemize}

To the best of our knowledge, this is the first work which successfully combines queueing theory and stochastic geometry for the optimization of AoI in wireless networks.
In addition, several mathematical results are also new:
$i$) the characterization of the scheduling policy which exploits spatial information from any stopping set,
$ii$) closed-form expressions for the transmission success probabilities,
and $iii$) the analytical expression for peak AoI in the context of wireless networks.

The remainder of the paper is organized as follows. We introduce the system model in Section II. There, we describe the quasi-static networks of interest, in which transmitters learn how to incorporate their location information in the scheduling policy design. In Section III, we focus on the stopping set based distributed algorithm design, which leads to a locally adaptive scheduling policy. We show that nodes can compute the channel access probability as solutions to certain fixed point equations. Section IV contains the analytical performance results. We show the numerical results in Section V to quantify the benefit of using local information to design a scheduling policy and minimize the information age. Finally, we conclude the paper in Section VI.

%%%%%%%%%%%%%%%%%%%%%%%%%%%%%%%%%%%%%%%%%%%%%%%%%%%%
\begin{table}
\caption{Notation Summary
%\mynote{revise notation}
} \label{table:notation}
\begin{center}
%\rowcolors{2}%{green!10!yellow}{}
%{cyan!15!}{}
\renewcommand{\arraystretch}{1.3}
%\begin{tabular}{c  p{6.0cm} }
\begin{tabular}{c  p{5.5cm} }
\hline
 {\bf Notation} & {\hspace{2.5cm}}{\bf Definition}
\\
%\midrule
\hline
$\tilde{\Phi}$; $\lambda$ & PPP modeling the location of transmitters; transmitter spatial deployment density \\
$\bar{\Phi}$; $\lambda$ & PPP modeling the location of receivers; receiver spatial deployment density \\
$P_{\mathrm{tx}}$; $\alpha$ &  Transmit power; path loss exponent \\
$\xi$; $T$ & Packet arrival rate; SINR decoding threshold \\
$\mu_{0, t}^\Phi$ & Transmission success probability of node $0$ at time slot $t$, conditioned on the point process $\Phi$ \\
$a_j^\Phi$ & Queue non-empty probability at transmitter node $j$, conditioned on the point process $\Phi$ \\
$A^{\mathrm p}$ & Network peak age of information \\
$S$  & Locally stopping set, contains all the observable information from any transmitter \\
$\eta_{S}(\theta_{X_i}\!\tilde{\Phi}, \theta_{X_i}\!\bar{\Phi} )$ & Channel access probability of transmitter $i$, constructed based on the stopping set $S$ and the point processes $\tilde{\Phi}$ and $\bar{\Phi}$ \\
\hline
\end{tabular}
\end{center}\vspace{-0.63cm}
\end{table}%
% ================================== %
%            System Model            %
% ================================== %
\section{System Model}
In this section, we provide a general introduction to the network topology, the traffic profile, as well as the concept of peak AoI and stopping sets. The main notations used throughout the paper are summarized in Table~I.

\subsection{Network Structure}
We model the wireless network as a set of transmitters and their corresponding receivers, all located in the Euclidean plane. The transmitting nodes are scattered according to a homogeneous Poisson point process (PPP) $\tilde{\Phi}$ of spatial density $\lambda$. Each transmitter located at $X_i \in \tilde{\Phi}$ has a dedicated receiver, whose location $y_i$ is at distance $r$ in a random orientation.
According to the displacement theorem \cite{BacBla:09}, the location set $\bar{\Phi} = \{y_i\}_{i=0}^\infty$ also forms a homogeneous PPP with spatial density $\lambda$.
We segment the time into slots with the duration of each slot equal to the time to transmit a single packet.
The packet arrivals at each transmitter are modeled as independent and identically distributed (i.i.d.) Bernoulli processes with parameter $\xi$.
All incoming packets are stored in a single-server queue with infinite capacity under the FCFS discipline\footnote{Note that the framework developed in this paper can also be extended to account for other transmission protocols, e.g., the LCFS with preemption \cite{YanXuWan:19}}.
During each time slot, the queue-nonempty transmitter will initiate a channel access attempt according to its scheduling protocol, and send out one packet upon approval. The transmission succeeds if the signal-to-interference-plus-noise ratio (SINR) at the corresponding receiver exceeds a predefined threshold.
A packet is removed from the buffer when its reception is acknowledged by the receiver through an ACK feedback. If the packet is not correctly decoded, the receiver sends a NACK message and the packet is retransmitted. We assume the ACK/NACK transmission is instantaneous and error-free, as commonly done in the literature \cite{TalKarMod:18}.
In order to investigate the time-domain evolution, we limit the mobility of transceivers by considering a static network, i.e., the locations of transmitters and receivers remain unchanged in all the time slots.

We assume that each transmitter uses unit transmission power $P_{\mathrm{tx}}$\footnote{ We unify the transmit power to keep the analysis tractable, it shall be noted that the results from this paper can be extended to account for power control via similar approach as in \cite{GhaElsBad:17}. }. The channel is subjected to both Rayleigh fading, which varies independently across time slots, and path-loss that follows power law attenuation.
Moreover, the receiver is also subjected to white Gaussian thermal noise with variance $\sigma^2$. By applying Slivnyak's theorem \cite{BacBla:09}, it is sufficient to focus on a \textit{typical} receiver located at the origin, with its tagged transmitter at $X_0$. Thus, when the tagged transmitter sends out a packet during slot $t$, the corresponding SINR received at the typical node can be written as
\begin{align}
\mathrm{SINR}_{0,t} = \frac{P_{\mathrm{tx}} H_{00} r^{-\alpha} }{ \sum_{ j \neq 0 } P_{\mathrm{tx}} H_{j0} \zeta_{j,t} \nu_{j,t} \Vert X_j \Vert^{-\alpha} + \sigma^2 }
\end{align}
where $\alpha$ denotes the path loss exponent, $H_{ji} \sim \exp(1)$ is the channel fading from transmitter $j$ to receiver $i$, $\zeta_{j,t} \in \{ 0, 1 \}$ is an indicator showing whether the buffer of node $j$ is empty ($\zeta_{j,t}=0$) or not ($\zeta_{j,t}=1$), and $\nu_{j,t} \in \{ 0, 1 \}$ represents the scheduling decision of node $j$, where it is set to 1 upon assuming transmission approval and 0 otherwise.

{\remark{\textit{From an engineering point of view, our system model is motivated by the emerging interest in applications like Device-to-Device (D2D) networking, mobile crowd sourcing, and Internet-of-Things (IoT), which do not require a centralized infrastructure, e.g., bases stations or access points, to conduct communications. Note that the framework can be extended to consider other channel models such as the multiple access channels (multiple transmitters and one receiver) \cite{GhaElsBad:17} or broadcast channels (one transmitter and multiple receivers) \cite{YanWanQue:18}, or even scenarios with multi-hop transmissions \cite{BacBlaMuh:06}.}	
}}

\begin{figure}[t!]
  \centering{}

    {\includegraphics[width=0.95\columnwidth]{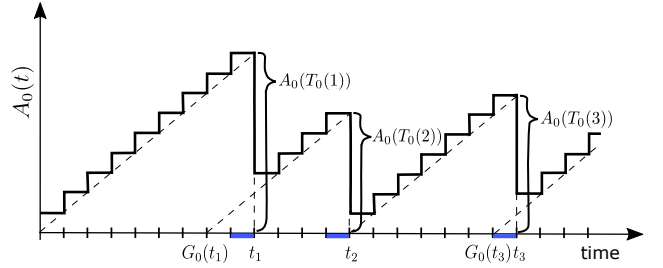}}

  \caption{ An example of the time evolution of age at a typical link. The time instance $T_0(i)$ denotes the moment when the $i$-th packet is successfully delivered, and $G(T_0(i))$ is the generation moment of the packet to be transmitted in the time slot following $T_0(i)$.
  The age is set to be $T_0(i) - G(T_0(i)) + 1$. }
  \label{fig:AoIMod_V1}
\end{figure}
\subsection{Age of Information}
Without loss of generality, we denote the communication link between the transmitter-receiver pair located at $(X_0, y_0)$ as \textit{typical}.
Then, as illustrated in Figure~\ref{fig:AoIMod_V1}, the AoI $A_0(t)$ over the typical link grows linearly in the absence of successful communication, and, when the transmission is successful, reduces to the time elapsed since the generation of the delivered packet. To make the statement more precise, we formalize the evolution of $A_0(t)$ via the following expression:
\begin{align*}
A_0(t \!+\! 1) =
\left\{
       \begin{array}{ll}
         \!\!  A_0(t)  +   1, \quad \quad ~~   \text{if transmission fails}, \\
         \!\!  t  -  G_0(t)  +  1, \quad \text{otherwise}
       \end{array}
\right.
\end{align*}
where $G_0(t)$ is the generation time of the packet delivered over the typical link at time $t$.

In this paper, we use the \textit{peak AoI} as our metric to evaluate the age performance across a wireless network\footnote{ We focus on the peak AoI because it is often the maximum status information delay that determines the performance loss in many wireless systems \cite{Cha:12}. Besides, compared to the average AoI, the peak AoI often possesses a simpler form and hence facilitates many low-complexity designs.
Nevertheless, the theoretical framework developed in this paper can be extended to analyze the average AoI \cite{CosCodEph:16}, given rise to a more involved computation.
% We focus on the peak AoI because it is often the maximum status information delay that determines the performance loss in many wireless systems \cite{Cha:12}. Nevertheless, the developed framework can be extended to analyze the average AoI \cite{CosCodEph:16}, given rise to a more involved computation.
}. Formally, the peak AoI at one generic link $j$ is defined as
\begin{align}
A^{ \mathrm{p} }_j = \limsup\limits_{ N \rightarrow \infty } \frac{ \sum_{n=1}^N A_j( T_j(n) ) }{N},
\end{align}
where $T_j(n)$ is the time slot at which the $n$-th packet from link $j$ is successfully delivered. We can extend this concept to a network scale and define the \textit{network} peak AoI as follows:
\begin{align*}
A^{\mathrm{p}} &= \limsup_{R \rightarrow \infty} \frac{ \sum_{ X_j \in \tilde{\Phi} \cap B(0,R) } A^{ \mathrm{p} }_j }{ \sum_{ X_j \in \tilde{\Phi} } \chi_{ \{ X_j \in B(0,R) \} }  }\\
&\stackrel{(a)}{=}   \mathbb{E}^0 \Big[ \limsup_{ N \rightarrow \infty } \frac{1}{N} \sum_{n=1}^N A_0( T_0(n) )  \Big]
\end{align*}
where $B(0,R)$ denotes a disk centered at the origin with radius $R$, $\chi_{E}$ is an indicator function which takes value 1 if event $E$ occurs and 0 otherwise, and $(a)$ follows from Campbell's theorem \cite{BacBla:09}. The notion $\mathbb{E}^0[\cdot]$ indicates the expectation is taken with respect to the Palm distribution $\mathbb{P}^0$ of the stationary point process, where under $\mathbb{P}^0$ almost surely there is a node located at the origin \cite{BacBla:09}.

\subsection{Stopping Sets and Scheduling Policy}
In a wireless network, as all transmitters are intertwined through the interference they cause to each other, it is important to have an effective protocol that schedules the appropriate channel access state for each node.
Inspired by the fact that knowledge from local activities can be utilized to improve the overall network performance, we incorporate such local information in the design of the scheduling policy.

\begin{figure}[t!]
  \centering{}

    {\includegraphics[width=0.95\columnwidth]{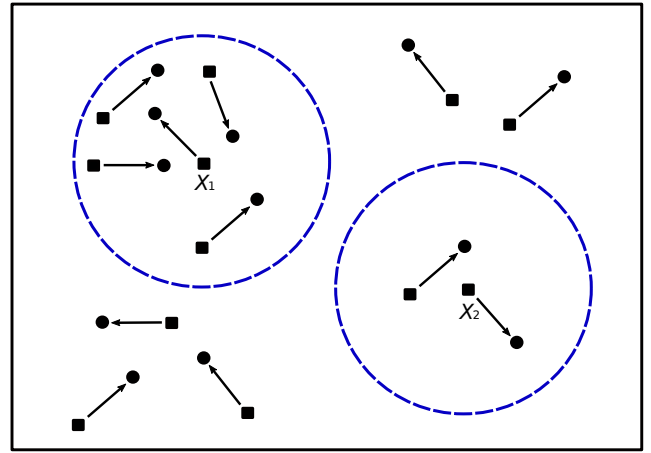}}

  \caption{ Illustration of a Poisson bipolar network with stopping sets being disks with constant radii, where black squares and dots are the transmitters and receivers, respectively, and the circles with dashed lines are two exemplary stopping sets centered at $X_1$ and $X_2$. }
  \label{fig:StpSet_V1}
\end{figure}

For a generic transmitter, note that it can only obtain the information about its geometry vicinity, we thus encapsulate such local knowledge by the notion of \textit{stopping set} $S = S(\tilde{\Phi}, \bar{\Phi})$ \cite{BacBla:09,BacBlaSin:14}.
More precisely, the stopping set is a random element taking each realization from the Borel sets in $\mathbb{R}^2$ such that for any observation window $A$, one can determine whether $S = S(\tilde{\Phi}, \bar{\Phi}) \subset A$.
This concept enables us to model the region in which the information of nodes, including their locations, are known to a typical observer.
In particular, depending on the scenarios under consideration, stopping sets can take various forms.
For instance, if the transmitters have unified sensing power, the observation region at each node will be a disk with a constant radius and the stopping set takes a deterministic form. When the transmitters are empowered with heterogeneous
sensing capabilities, each node may want to obtain information up to
the $p$-th nearest neighbor, in which case the observation region varies across different nodes and the stopping set takes a random shape. Aside from disks, the stopping set can have more complicated formats, e.g., a hexagon under clustering regulation or different orders of Voronoi cells in the context of cellular
networks \cite{YanGerQue:17}, depending on the specific task under consideration.
An illustration of deterministic stopping sets in a Poisson bipolar network is given in Fig.~\ref{fig:StpSet_V1}. Note that different transmitters, e.g., the nodes located at $X_1$ and $X_2$, can have various local observations.

To generalize the concept network wide, we further introduce a shifting operation, denoted by $\theta_x$ and performs on $\tilde{\Phi}$ and $\bar{\Phi}$, which translates all the network nodes by the vector $-x$, i.e., $\theta_{x}\{X_i\} = \{ X_i - x \}$ and similarly for the receivers.
Extending this operator to a Borel $A \subset \mathbb{R}^2$, we have $\theta_x(A) = \{ a - x: a \in A \}$.
Using the shifting operator and the stopping set, we define the set of candidated scheduling policies to be the ones that take into the local knowledge at every source node and output a channel access probability for it.
More precisely, given a stopping set $S = S(\tilde{\Phi}, \bar{\Phi})$, the scheduling policy at the typical transmitter can be defined as follows:
\begin{align} \label{equ:eta_S0}
\gamma^\Phi_0 &= \eta_{\mathrm{S}}(\tilde{\Phi}, \bar{\Phi})
\nonumber\\
&= \eta_{\mathrm{S}}\!\left( \tilde{\Phi} \cap S, \bar{\Phi} \cap S \right),
\end{align}
where $\eta_{\mathrm{S}}(\cdot)$ is a measurable function whose argument is the network geometry ($\tilde{\Phi}$, $\bar{\Phi}$) and has value in $[0, 1]$. For node $i$ located at $X_i$, its scheduling policy can be obtained by applying the shifting operator $\theta_{X_i}$ to \eqref{equ:eta_S0}, which resulted in $\gamma_i^\Phi = \eta_{\mathrm{S}}(\theta_{X_i}\!\tilde{\Phi}, \theta_{X_i}\!\bar{\Phi} )$.
To this end, a feasible scheduling policy is \textit{translation invariant} which allows us to devise it by only focusing on the typical node.
Note that to apply such a policy and evaluate the channel access probability $\eta_{\mathrm{S}}( \theta_{X_i}\! \tilde{\Phi}, \theta_{X_i}\! \bar{\Phi} )$, node $i$ needs to obtain only local knowledge about the other nodes in the stopping set $S_i = {S}(\theta_{X_i}\! \tilde{\Phi}, \theta_{X_i}\! \bar{\Phi})$, which can be obtained via, e.g., the methods in \cite{BacBlaSin:14}.
In this regard, the scheme can run without a central controller and is thus decentralized.

{\remark{\textit{By leveraging the notion of stopping sets, our framework is able to provide a unified approach to account for various types of local information.}}}

{\remark{\textit{When each source node in this network is able to obtain the location information of its nearest neighbors, it is also possible to devise a distributed power control mechanism to reduce the AoI in a similar spirit to \cite{RamBac:19}. }	
}}

% ================================== %
%            Scheme Design           %
% ================================== %
\section{Scheduling Policy Design}
We now present the main structural result of this paper. Specifically, by leveraging the spatial information contained in locally stopping sets, we design a low-complexity scheduling policy to reduce the peak AoI of a wireless network.
\subsection{Preliminaries}
The radio interface between any transmitter-receiver pair $i$ can be modeled as a Geo/G/1 queue where the departure rate varies according to the link throughput.
In the steady state, the link throughput, or equivalently service rate, is determined by both the scheduled channel access probability, i.e., how frequent a transmitter with non-empty buffer can access the channel, and the transmission success probability. Particularly, conditioned on the realization of the point process  $\Phi \triangleq \tilde{\Phi} \cup \bar{\Phi}$, the transmission success probability, $\mu^\Phi_i$, is given by \cite{ZhoQueGe:16,YanQue:19}
% \footnote{ In the following, we will drop the time index $t$ from the subscript as we are dealing with the situation under steady state. }
\begin{align} \label{equ:TXSucProb}
\mu_i^\Phi = \mathbb{P}\left( \mathrm{SINR}_i > T | \Phi  \right)
\end{align}
where $T$ is the decoding threshold.
We first average out the randomness from channel fading and derive a conditional form of the transmission success probability:
\begin{lemma}
\textit{Conditioned on the spatial realization $\Phi$, the transmission success probability at the typical link during time slot $t$ is given by
	\begin{align} \label{equ:CondThrPut}
	\mu_{0,t}^\Phi = e^{-\frac{ T r^\alpha }{\rho}}  \mathbb{E}\bigg[ \prod_{j \neq 0} \Big(  \frac{ 1 }{ 1 + \zeta_{j,t} \nu_{j,t} / \mathcal{D}_{j0} } \Big) \Big\vert \Phi \bigg]
	\end{align}
	where $\rho = P_\mathrm{tx}/\sigma^2$ and $\mathcal{D}_{ij} = \Vert X_i - y_j \Vert^\alpha / T r^\alpha$.
}
\end{lemma}
\begin{IEEEproof}
Conditioned on the spatial realization of all the transceiver locations, the transmission success probability can be derived as follows:
\begin{align}
&\mathbb{P}\!\left( \mathrm{SINR}_{0,t} \!>\! T | \Phi  \right)
\nonumber\\
=& \mathbb{P} \bigg(  \frac{ P_{ \mathrm{tx} } H_{00} r^{-\alpha} }{ \sum_{ j \neq 0 } { P_{ \mathrm{tx} } H_{j0} \zeta_{j,t} \nu_{j,t} }{ \Vert X_j \Vert^{-\alpha} } \!+\! \sigma^2 }  >  T \,  \Big| \, \Phi  \bigg)
\nonumber\\
=& \mathbb{P} \bigg(  { H_{00} }  > { T r^\alpha } \Big( \sum_{ j \neq 0 } \frac{ H_{j0} \zeta_{j,t} \nu_{j,t} }{ \Vert X_j \Vert^\alpha } \!+\! \frac{1}{\rho} \Big) \,  \Big| \, \Phi  \bigg)
\nonumber\\
\stackrel{(a)}{=} & \mathbb{E}\bigg[ e^{-\frac{ T r^\alpha }{\rho}} \prod_{ j \neq 0 } \exp\! \Big(\! - T r^\alpha \frac{ H_{j0} \zeta_{j,t} \nu_{j,t} }{ \Vert X_j \Vert^\alpha } \Big)  \Big| \Phi \bigg],
\end{align}
where $(a)$ follows by leveraging the Rayleigh distribution of the channel gain, i.e., $H_{00} \sim \exp(1)$, and the result then follows by noticing that the random variables $\{ H_{j0}, j = 1, 2, ... \}$ are i.i.d. and exponentially distributed as $H_{j0} \sim \exp(1)$.
\end{IEEEproof}
Lemma~1 presents a general result for wireless queueing networks. Note that because the spectrum is shared amongst the transmitters, their active states, $\zeta_{j,t}$, are by nature correlated, regardless of the transmission decisions, $\nu_{j,t}$, are made independently or not, due to the interference they cause to each other. These phenomena are commonly known as the \textit{spatially interacting queues}, and, at the current stage, there is no comprehensive theory to characterize it \cite{SanBacFos:19}. Therefore, we need to opt for a few approximations in the sequel to trade for usable results.

Next, by conditioning on the realization of the point process $\Phi$, the communication between a typical transceiver pair can be regarded as a Geo/Geo/1 queue where the service rate is given by $\gamma_0^\Phi \mu_{0,t}^\Phi$.
As such, using tools from queueing theory, we arrive at a conditional form of the peak AoI.
\begin{lemma}\label{lma:Cond_AoI}
\textit{In the steady state, conditioned on the spatial realization $\Phi$, the peak AoI at a typical link is given as
	\begin{align} \label{equ:CondPeakAoI}
  \mathbb{E}^0\!\left[ A^{\mathrm{p}} | \Phi \right] = \left \{ \!\!\!
\begin{tabular}{cc}
$ \frac{1}{\xi} +\!  \frac{ 1 - \xi }{ \gamma_0^\Phi \mu_0^\Phi - \xi }$, & if $\, \gamma_0^\Phi \mu_0^\Phi  > \xi$,   \\
+$\infty$, &  if $\, \gamma_0^\Phi \mu_0^\Phi \leq \xi$
\end{tabular}
\right.
	\end{align}
where $\mu_0^\Phi = \lim_{ t \rightarrow \infty } \mu_{0,t}^\Phi$.
}
\end{lemma}
\begin{IEEEproof}
See Appendix~H of \cite{TalSerEyt:18} for a detailed proof.
\end{IEEEproof}

From \eqref{equ:CondPeakAoI}, it is obvious that the principle of optimizing information freshness consists in maximizing the link throughput.
In order to achieve this goal, a policy that schedules the channel access at each node by jointly balancing the radio resource utility and mutual interference is essential.
In the following, we formulate a stochastic decision problem to find the scheduling policy that accomplishes this task.
\subsection{Locally Adaptive Scheduling Policy}
\subsubsection{Design} Let a stopping set $S = S(\tilde{\Phi}, \bar{\Phi})$ be given. Using Lemma~\ref{lma:Cond_AoI}, the design of scheduling policy can be written as:
\begin{align} \label{prbm:PeakAoI}
& \min_{ \eta_{\mathrm{S}} } ~~~~ \mathbb{E}^0_\Phi\! \left[ \frac{ 1 - \xi }{ \gamma_0^\Phi \mu_0^\Phi - \xi } \right] + \frac{1}{\xi}\\ \label{equ:Rst_SpSt}
& ~~ \mathrm{s.t.} \quad~ 0 \leq \gamma_i^\Phi = \eta_{\mathrm{S}}\!\left( \theta_{X_i} \! \tilde{\Phi}, \theta_{X_i}\! \bar{\Phi} \right)  \leq 1, ~~ \forall i \\ \label{equ:Rst_Stb}
& \qquad \quad ~ \xi \leq \mathbb{E}^0_{\Phi}[ \eta_{\mathrm{S}} \mu_0^\Phi ].
\end{align}
It is worthwhile to point out that the design factor $\eta_{ \mathrm{S} }$ in \eqref{prbm:PeakAoI} is not a single parameter but instead a policy, which takes input the state information, i.e., the node's location and information observed from the corresponding stopping set, and as an output the channel access probability.
As such, the scheduling policy varies from node to node, which is stated in constraint \eqref{equ:Rst_SpSt}, because the local knowledge is location dependent.
Moreover, the queueing stability shall be guaranteed in the average sense, as shown in \eqref{equ:Rst_Stb}, according to Loynes' theorem \cite{Loy:62}.

However, according to \eqref{equ:CondThrPut}, we find the optimization problem \eqref{prbm:PeakAoI} is hardly solvable because $\mu_0^\Phi$ does not even possess an analytical expression.
For that reason, we leverage the dominant system \cite{BorMcDPro:12}, in which every transmitter keeps sending out packets in each time slot (if one transmitter has an empty buffer at any given time slot, it sends out a dummy packet), for an approximation.
Since each node is backlogged in the dominant system, that allows us to unpack the interaction amongst queues and derive a closed-form expression for the conditional transmission success probability:
\begin{lemma} \label{lma:CndtSucProb_Dmnt}
\textit{Given stopping set $S$ and conditioned on the spatial realization $\Phi$, the transmission success probability at the typical link under the dominant system is given by
	\begin{align} \label{equ:Dmnt_SucProb}
	\hat{\mu}_{0}^\Phi = e^{-\frac{ T r^\alpha }{\rho}}  \prod_{j \neq 0} \Big(  1 -  \frac{ \gamma^\Phi_j }{ 1 +  \mathcal{D}_{j0} } \Big)
	\end{align}
	where $\rho = P_\mathrm{tx}/\sigma^2$, $\gamma_j^\Phi = \eta_{\mathrm{S}}( \theta_{X_j} \! \tilde{\Phi}, \theta_{X_j}\! \bar{\Phi})$, and $\mathcal{D}_{ij} = \Vert X_i - y_j \Vert^\alpha / T r^\alpha$.}
\end{lemma}
\begin{IEEEproof}
See Appendix~\ref{apx:CndtSucProb_Dmnt}.
\end{IEEEproof}

Besides, because transmitters under the dominant sytem are subject to higher interference levels, the policies devised from a dominant sytem can also help us prepare for the ``worst case'' interference condition in the original system when all the transmitters are active.
As such, instead of directly solving the original problem \eqref{prbm:PeakAoI}, we minimize the following alternative:
\begin{align} \label{prbm:Dom_PeakAoI}
& \min_{ \eta_{\mathrm{S}} } ~~~~ \mathbb{E}^0_\Phi\! \left[ \frac{ 1 - \xi }{ \gamma_0^\Phi \hat{\mu}_0^\Phi - \xi } \right] + \frac{1}{\xi}\\ \label{equ:Dom_Rst_SpSt}
& ~~ \mathrm{s.t.} \quad~ 0 \leq \gamma_i^\Phi = \eta_{\mathrm{S}}\!\left( \theta_{X_i} \! \tilde{\Phi}, \theta_{X_i}\! \bar{\Phi} \right) \leq 1, ~~ \forall i \\
& \qquad \quad ~ \xi \leq \mathbb{E}^0_{\Phi}[ \eta_{\mathrm{S}} \hat{\mu}_0^\Phi ]
\end{align}
where $\hat{\mu}_0^\Phi$ is given by \eqref{equ:Dmnt_SucProb}.
The design of the scheduling policy can now be cast into an optimization problem with explicit terms.
And that brings us to the main structural result of this paper.
\begin{theorem}\label{thm:DmSym_PeakAoI}\textit{For all given stopping sets $S = S(\tilde{\Phi}, \bar{\Phi})$, the solution to the optimization problem in \eqref{prbm:Dom_PeakAoI} is given by the unique solution of the following fixed point equation:
	\begin{align}\label{equ:OptSln}
	\frac{1}{\eta_{\mathrm{S}}} -\!\!\!\!\! \sum_{ \substack{ j \neq 0, y_j \in S } } \frac{1}{ 1  \!+\! \mathcal{D}_{0j} \!-\! \eta_{\mathrm{S}} } -\!\! \int_{\mathbb{R}^2 \setminus S }\! \frac{ \lambda  dz }{ 1 \!+\! \Vert z \Vert^\alpha \! / T r^\alpha } = 0
	\end{align}
	if the following condition holds
	\begin{align} \label{equ:CndOptl}
	\sum_{ \substack{ j \neq 0, y_j \in S } }  \frac{1}{ \mathcal{D}_{0j} } +\! \int_{ \mathbb{R}^2 \setminus S }\! \frac{ \lambda dz }{ 1 \!+\! \Vert z \Vert^\alpha \! / T r^\alpha  } > 1.
	\end{align}
	Otherwise, $\eta_{\mathrm{S}} = 1$.
	% Moreover, for any other $\eta_{\mathrm{S}}'$ that solves the problem in \eqref{prbm:Dom_PeakAoI}, we have $\eta_{\mathrm{S}}'(\tilde{\Phi}, \bar{\Phi}) = \eta_{\mathrm{S}}(\tilde{\Phi}, \bar{\Phi})$ for almost all realizations of $(\tilde{\Phi}, \bar{\Phi})$.
}
\end{theorem}
\begin{IEEEproof}
See Appendix~\ref{apx:OptCtrl_PeakAoI}.
\end{IEEEproof}

The above theorem gives an explicit way to construct the scheduling policy at the typical node, i.e., $\gamma_0^\Phi = \eta_{ \mathrm{S} }(\theta_{ X_0 } \tilde{\Phi}, \theta_{ X_0 } \bar{\Phi} )$.
In regard to a generic node $i$, the corresponding policy can be attained by shifting the origin of the point process $\Phi$ to $X_i$ and then apply the same approach as per Theorem~1, which gives $\gamma_i^\Phi = \eta_{ \mathrm{S} }(\theta_{ X_i } \tilde{\Phi}, \theta_{ X_i } \bar{\Phi} )$. As such, every source node only need to identify and record the transmitting neighbors located inside its observation window, i.e., the stopping set $S$, and solve for the channel access probability via a fixed point equation, which has very low complexity.
Essentially, such a scheduling policy is a spatial version of ALOHA, where the channel access probability at each transmitter is conceived based on the local topology information. Different from the interference graphs \cite{talak2018optimizing}, which mute simultaneous transmissions from nodes in proximity locations, the proposed approach allows neighboring transmitters to initiate the radio channel access attempts at the same time while communally controlling the mutual interference with a location-dependent probability and thus achieves better spectrum reuse.

It is also worthwhile to mention that condition \eqref{equ:CndOptl} implies the typical transmitter will only opt for an opportunistic channel access when the following holds:
\begin{align*}
T &> \frac{ { r^{-\alpha} } }{ \sum\limits_{ j \neq 0, y_j \in S } \!\!\!\!\! { \Vert y_j \Vert^{-\alpha} } \!+\! \int_{ \mathbb{R}^2 \setminus S } \frac{ \lambda dz }{ \Vert z \Vert^\alpha + r^\alpha } } \\
&> \frac{ { r^{ - \alpha} } }{ \sum\limits_{ j \neq 0, y_j \in S } \!\!\!\!\! { \Vert y_j \Vert^{-\alpha} } \!+\! \int_{ \mathbb{R}^2 \setminus S } \frac{ \lambda dz }{ \Vert z \Vert^\alpha  } } \\
& = \frac{ \mathbb{E} \big[ P_{\mathrm{tx}} H_{00} r^{-\alpha} \big] }{ \mathbb{E}\big[ \sum_{ j \neq 0 } P_{\mathrm{tx}} H_{0j} \Vert X_0 - y_j \Vert^{-\alpha} | S \big] },
\end{align*}
namely the ratio between the average signal power and interference generated by the transmitter is smaller than the decoding threshold.
In other words, a given source node will reduce the channel access frequency when its own transmission may cause potential transmission failure to the neighbors.
Such an observation also coincides with the intuition that transmitters located close to each other can cause severe mutual interference and hence need to be scheduled for more stringent channel access, while the ones located far away from their neighbors can access the radio channel more frequently.

\subsubsection{Examples} Below, we illustrate the proposed scheme via a few examples to better understand the results of the theorem. To keep the results intuitive, we limit the example stopping sets to be disk-based, but note that the framework is quite versatile and can accommodate more general situations, e.g., with stopping sets being the irregular extended cells \cite{YanGerQue:17}.

$a) ~S = \emptyset$: When transmitters have no topological information about their neighbors, the scheduling policy shall be assigned as a universal constant. Using results from Theorem~\ref{thm:DmSym_PeakAoI}, we have $\eta_{ \mathrm{S} }(X_i)=1, \forall i \in \mathbb{N}$.
This can also be recognized by the fact that the average link throughput
\begin{align*}
\mathbb{E}^0[ \gamma_0 \hat{\mu}_0 ] = \eta_{ \mathrm{S} } \exp\big( - \int_0^\infty \!\!\! \frac{ 2 \pi \lambda \eta_{ \mathrm{S} } v dv }{ 1 + v^\alpha / Tr^\alpha } \big)
\end{align*}
monotonically increases with respect to $\eta_{ \mathrm{S} }$. Thus, the best strategy is to set $\eta_{\mathrm{S}} = 1$ at every transmitter.

$b)~ S = B(0,\Vert y_\mathrm{c} \Vert)$: Here, for a generic transmitter $j$, $y_{\mathrm{c}}$ denotes the nearest receiver that node $j$ generates interference to.
This corresponds to the scenario where $S$ is a random stopping set.
By solving \eqref{equ:OptSln}, the scheduling policy takes the following form:
\begin{align*}
{\eta_{\mathrm{S}}} \!=\! \frac{ 1 }{ \mathcal{M}( \Vert y_\mathrm{c} \Vert ) } \!+\! \frac{ \Vert y_\mathrm{c} \Vert^\alpha \!\!+\! T r^\alpha }{ 2 T r^\alpha } \!-\! \sqrt{ \frac{ T r^\alpha \!+\! \Vert y_\mathrm{c} \Vert^\alpha  }{ 4 T r^\alpha } \!+\! \frac{ 1 }{ \mathcal{M}( \Vert y_\mathrm{c} \Vert )^2 } },
\end{align*}
where $\mathcal{M}( \Vert y_\mathrm{c} \Vert )$ is given as
\begin{align*}
\mathcal{M}( \Vert y_\mathrm{c} \Vert ) = \int_{ \Vert y_\mathrm{c} \Vert }^\infty \frac{ 2 \pi \lambda v dv }{ 1 + v^\alpha / T r^\alpha }.
\end{align*}

$c)~ S = B(0,R)$: In this example, each transmitter knows the location of receivers in a disk of radius $R$ centered at its location, with $R>0$. Note that such stopping set is a deterministic set. The proposed policy can thus be attained by solving
\begin{align*}
\frac{1}{ \eta_{\mathrm{S}} } =\!\!\!\!\! \sum_{ 0 < \Vert y_j \Vert \leq R } \frac{1}{ \frac{ \Vert y_j \Vert^\alpha }{ T r^\alpha } \!+\! 1 \!-\! \eta_{\mathrm{S}}  } +  2 \pi \! \int_{R}^\infty \! \frac{  \lambda v dv }{ 1 \!+\! { v^\alpha } / { T r^\alpha } }.
\end{align*}

% \subsubsection{Implementation}
% %Purely understanding the theories are not enough, we would like to know how they can be applied in practice.
% We now mention a few remarks regarding how to implement the scheduling policy.
% According to Theorem~\ref{thm:DmSym_PeakAoI}, the implementation of the proposed scheme only needs each transmitter to $i$) identify and record the transmitting neighbors located inside the stopping set $S$, and $ii$) estimate the network spatial density $\lambda$.
% These parameters can be obtained via, e.g., the methods in \cite{BacBlaSin:14}, and each node can then easily solve for its channel access probability.
% To this end, the proposed scheme has a very low density and can be efficiently implemented.
% The performance of the proposed protocol will be quantitatively analyzed in Section IV and numerically evaluated in Section V.

The effectiveness of the proposed protocol will be amply illustrated by numerical examples in Section V. Before that, we would like to present a few quantitative results to assess the performance of the developed scheme.

% ================================== %
%              Analysis              %
% ================================== %
\section{Performance Analysis}
In this section, we derive analytical expressions to characterize the stochastic behavior of several network statistics when the proposed scheme is employed.
Specifically, we analyze the distribution of channel access probabilities resulted from our proposed scheduling policy, the conditional transmission success probability, and the mean value of Peak AoI. For better readability, most proofs and mathematical derivations have been relegated to the Appendix.
\subsection{Distribution of the Scheduling Policy}
Due to randomness in the node locations, information observed from the stopping sets varies from different transmitters and so does the scheduling policy.
Hence, the channel access probability, $\gamma_i^\Phi = \eta_{\mathrm{S}}( \theta_{X_i} \! \tilde{\Phi}, \theta_{X_i}\! \bar{\Phi})$, at a generic node $i$ is a random variable whose behavior can only be captured by the distribution.
In the following, we derive the analytical expressions for the distribution of such a quantity\footnote{From now on, we use $\eta_S$ to denote both the scheduling policy and the resultant channel access probability at a generic node when that does not cause ambiguity.}.
To maintain the computational complexity at a reasonable level, we constrain the stopping sets to take deterministic forms, though a generalization to random sets is straightforward (cf. Remark 3).
\begin{theorem}\label{thm:etaS_distribution}\textit{Given a deterministic set $S$ and $0 < \kappa <1$, the complementary cumulative distribution function (CCDF) of the channel access probabilities resulted from the proposed scheduling policy has the following expression:
    \begin{align} \label{equ:SchPlc_Dist}
        \mathbb{P} \left( \eta_{\mathrm{S}} > \kappa \right)  = \frac{1}{ 2 \pi } \!\! \int_{-\infty}^{+\infty} \!\!\!\!\!\!\!\! \mathcal{L}_{\mathcal{U}(\kappa, S)} (j \omega) \, \frac{ e^{j \omega ( 1 - \mathcal{V}( \kappa, S ) ) } - 1 }{ j \omega } d \omega,
    \end{align}
    and when $\kappa = 1$, the following holds:
	\begin{align}\label{equ:eta_one}
	\mathbb{P} \left( \eta_{\mathrm{S}} = 1 \right) = \mathbb{P}\left( \, \mathcal{U}( 1, S ) < 1 - \mathcal{V}(1, S) \,  \right),
	\end{align}
    where $j=\sqrt{-1}$, the quantities $\mathcal{U}( \kappa, S )$ and $\mathcal{V}(\kappa, S)$ are respectively defined as
	\begin{align} \label{equ:UkS}
	\mathcal{U}( \kappa, S ) &= \sum_{ y_i \in \bar{\Phi} } \frac{ \kappa \cdot \chi_{ \{ y_i \in S \} } }{ \frac{ \Vert y_i \Vert^\alpha }{ T r^\alpha } + 1 - \kappa },\\ \label{equ:VkS}
	\mathcal{V}(k,S) &= \int_{ \mathbb{R}^2 \setminus S } \frac{ \lambda \kappa  dz }{ 1 + \Vert z \Vert^\alpha \! / T r^\alpha },
	\end{align}
    whereas $\mathcal{L}_{\mathcal{U}(\kappa, S)} (s)$ is the Laplace transform of $\mathcal{U}( \kappa, S )$, given by
    \begin{align*}
    \mathcal{L}_{\mathcal{U}(\kappa, S)} (s) &= e^{  - \lambda  \int_{S} \left[ 1 - \exp\left( - \frac{s \kappa T r^\alpha}{ \Vert z \Vert^\alpha + ( 1 - \kappa ) T \! r^\alpha  } \right) \right] dz }.
    \end{align*}
}
\end{theorem}
\begin{IEEEproof}
See Appendix~\ref{apx:etaS_distribution}.
\end{IEEEproof}

The accuracy of this theorem will be verified in Fig.~\ref{fig:CDF_Schdl} in Section~V. A few important remarks are in order:

{\remark{\textit{Due to condition \eqref{equ:CndOptl} and the PPP model, the event $\eta_{ \mathrm{S} }=1$ takes a non-zero probability. In particular, we have $\mathbb{P} \left( \eta_{\mathrm{S}} = 1 \right) \rightarrow 1$ under the following scenarios:
	\begin{itemize}
	\item[$a$)]  $\lambda \rightarrow 0$, i.e., the network is sparse,
	\item[$b$)]	$|S| \rightarrow 0$, i.e., the observation region is very small,
	\item[$c$)] $r \rightarrow 0$, i.e., the received signal power is very strong.
	\end{itemize} } }}

{\remark{\textit{While the above result is derived under deterministic stopping sets, it also holds for random stopping sets by first conditioning on $S$ and then averaging on it. Furthermore, an approximation also follows by replacing $S$ by a disk $\bar{S} = B(0, \bar{R})$ where $\bar{R} = \sqrt{  \mathbb{E}[S] / \pi }$. } } }

 \newcounter{TempEqCnt}
 \setcounter{equation}{\value{equation}}
 \setcounter{equation}{21}
 \begin{figure*}[t!]
 \begin{align} \label{equ:Meta_Grl}
 F_{\mu}(u) \!\approx\! \frac{1}{2} - \!\! \int_0^\infty \!\!\!\!\! \mathrm{Im}\!\left\{ u^{-j \omega} \exp\!\bigg(\! - \frac{ j \omega T r^\alpha }{ \rho } - \lambda \pi r^2 T^\delta \sum_{k=1}^{\infty} \! \binom{j \omega}{k} \!\! \int\limits_0^\infty \! \frac{ (-1)^{ k+1 } }{ 1 \!+\! v^{ \frac{\alpha}{2} } } \Big[ \mathcal{Z}(v^{ \frac{\alpha}{2} })^k F_{ \mu }( \frac{\xi} {\mathcal{Z} ( v^{ \frac{\alpha}{2} } )} ) \!+\!\!\! \int_{ \xi / \mathcal{Z}( v^{ \frac{ \alpha }{ 2 } } ) }^1 \!\!\!\!\!\!\!\!\!\!\!\!\!\!\!\! \xi^k \!/ t^k F_{ \mu }(dt) \Big] dv \bigg) \! \right\} \! \frac{ d\omega }{ \pi \omega }
 \end{align}
 \setcounter{equation}{\value{equation}}{}
 \setcounter{equation}{22}
 \centering \rule[0pt]{18cm}{0.3pt}
 \end{figure*}
 \setcounter{equation}{\value{TempEqCnt}}
 \setcounter{equation}{20}
\subsection{ Transmission Success Probability and Stable Condition }
We next deal with the distribution of the conditional transmission success probability, $\mu^\Phi_0$, as defined in \eqref{equ:TXSucProb}\footnote{ In the following, we will drop the time index $t$ from the subscript as we are dealing with the situation under steady state.}.
It is worthwhile to mention that although we leverage the dominant system in Section 3 to devise the scheduling policy, the analysis presented in this section is conducted upon the original system which does not assume full buffer at the transmitters.

To begin with, because the scheduling policy is constructed via stopping sets, as given by Theorem~\ref{thm:DmSym_PeakAoI}, it is inevitably correlated with the distances to the neighboring transmitters.
The following result captures such behavior.
\begin{lemma} \label{lma:CndSlct}\textit{Given the distance between the typical receiver and a generic transmitter $j$ as $\Vert X_j - y_0 \Vert = l$, the probability that transmitter $j$ can be scheduled for channel access is given by the following
	\begin{align} \label{equ:Cnd_SlctProb}
	&\mathbb{P}( \nu_j = 1 | \zeta_j = 1, \Vert X_j - y_0 \Vert = l ) = \mathcal{Z}(l^\alpha / T r^\alpha )
	\nonumber\\
	&= \!\! \int_0^1 \!\!\! \int_{-\infty}^{+\infty} \!\!\!\!\!\!\!\! \mathcal{L}_{\mathcal{U}(\kappa, S)} (j \omega) \, \frac{ e^{j \omega ( 1 - \mathcal{V}( \kappa, S ) - \frac{ \kappa }{ 1 - \kappa + l^\alpha \!/ T r^\alpha } ) } - 1 }{ j 2 \pi \omega } d \omega d \kappa.
	\end{align}
}
\end{lemma}
\begin{IEEEproof}
See Appendix~\ref{apx:CndSlct}.
\end{IEEEproof}

\begin{figure}[t!]
  \centering{}

    {\includegraphics[width=0.95\columnwidth]{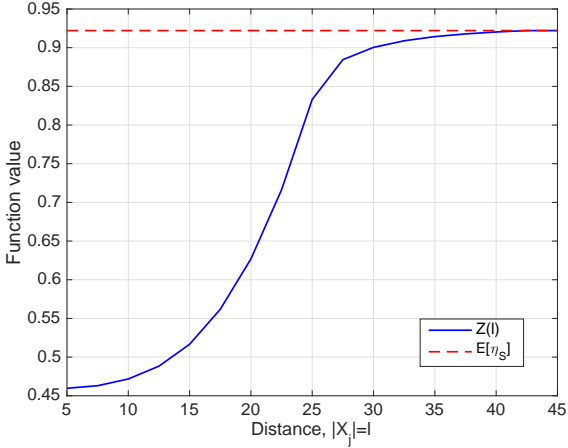}}

  \caption{ The value of function $\mathcal{Z}(\cdot)$ vs the distance to an interferer located at $X_j$: the system parameters are set as $\alpha = 3.8$, $T=0$ dB, $r=25$, $\lambda = 10^{-4}$, and $S=B(0,R)$ where $R=200$. }
  \label{fig:Z_Val}
\end{figure}

The function $\mathcal{Z}(\cdot)$ from the above characterizes the relationship between the channel access probability and the distance from the typical receiver
to a generic interfering transmitter.
As illustrated in Fig.~\ref{fig:Z_Val}, we can see that transmitters located close to the typical receiver will have a relatively small channel access probability in order to reduce the mutual interference, while such active probability asymptotically converges to a global constant, i.e., $\mathbb{E}[\eta_{\mathrm{S}}]$, as the distance increases.

Next, we derive the analytical expression for the distribution of the conditional transmission success probability $\mu_0^\Phi$. Note that there are still two issues associated with the analysis: $i$) due to interference, the active state, $\zeta_{j,t}$, at each transmitter $j$ is correlated with each other and varies over time, and $ii$) there may exist common transmitters seen by the same receiver from one time slot to another, which introduces temporal correlation.
The interaction among queues, together with the temporal correlation, incurs memory to the service process and can highly complicate the analysis.
Fortunately, such dependency of neighborhood queueing status becomes relatively weak at the macroscopic scale, which motivates the following assumption \cite{ChiElsCon:19}:
\begin{assumption}\textit{The temporal correlation of interference has a negligible effect on the transmission success probability. We thus assume the typical receiver sees almost independent interference at each time slot.
}
\end{assumption}
Assumption~1 is commonly known as the \textit{mean field approximation}, which ignores the spatiotemporal correlation between the buffer states of interfering nodes.
And based on this approximation, it is reasonable to assume that the service rate at all transmitters follows i.i.d. distribution in the steady state.

Armed with the above preparation, we are now ready to present the distribution of conditional transmission success probability.

\begin{theorem}\label{thm:SINR}\textit{In a wireless network with every transmitter adopting the scheduling policy as per Theorem~\ref{thm:DmSym_PeakAoI}, the CDF of the conditional transmission success probability at the typical link can be tightly approximated by equation \eqref{equ:Meta_Grl} at the top of this page, where $\delta = 2/ \alpha$, and $\mathrm{Im}\{ \cdot \}$ denotes the imaginary part of a complex variable.
}
\end{theorem}
\begin{IEEEproof}
See Appendix~\ref{apx:SINR_proof}.
\end{IEEEproof}

Notably, the effect of queueing interaction on the SINR is characterized by the fixed-point functional equation in~\eqref{equ:Meta_Grl}. Besides, \eqref{equ:Meta_Grl} can be solved via an iterative approach, and low-computational-complexity approximation is available to boost up the convergence speed \cite{YanQue:19}.

Motivated by the important role it plays in the network performance assessment, we further provide the first moment of $\mu^\Phi_0$, i.e., the transmission success probability \cite{AndBacGan:11} as an immediate byproduct of Theorem~\ref{thm:SINR}.
\begin{corollary}\textit{The transmission success probability is given by the solution of the following fixed-point equation
     \setcounter{equation}{\value{equation}}
     \setcounter{equation}{22}
    	\begin{align}
    	& \mathbb{P}(\mathrm{SINR}_0 > T) = \mathbb{E}[\mu_0^\Phi]
    	\nonumber\\
    	= &  \exp\!\Big( \! -\! \frac{ T r^\alpha }{ \rho } -\! \lambda \pi r^2 T^\delta \!\!\! \int_0^\infty \! \frac{ \min \! \big\{ \frac{ \xi }{ \mathbb{E}[\mu_0^\Phi] }, \mathcal{Z}( u^{ \frac{ \alpha }{ 2 } } ) \big\} }{ 1 + u^{ \frac{\alpha}{2} }} du \Big),
    	\end{align}
    	where $\mathcal{Z}(\cdot)$ is given in \eqref{equ:Cnd_SlctProb}.
     \setcounter{equation}{\value{equation}}{}
     \setcounter{equation}{23}
}
\end{corollary}
\begin{IEEEproof}
By taking an expectation of \eqref{equ:CondThrPut} with respect to the random measure $\Phi$, we have the transmission success probability given as
\begin{align} \label{equ:Emu0}
\mathbb{E}^0_\Phi[ \mu^\Phi_0 ] &= \mathbb{E}_\Phi\Big[ e^{ - \frac{ T r^\alpha }{ \rho } } \prod_{ j \neq 0 } \big( 1 - \frac{ a_j^\Phi \gamma_j^\Phi }{ 1 + \Vert X_i \Vert^\alpha / T r^\alpha } \big) \Big]
\nonumber\\
&= \exp \! \Big( \!  - \frac{ T r^\alpha }{ \rho } -  \lambda \pi \! \int_0^\infty \! \frac{ \mathbb{E}[ a_x \gamma_x ] 2 x dx }{ 1 + x^\alpha / T r^\alpha } \Big)
\end{align}
where $\mathbb{E}[ a_x \gamma_x ]$ is the active probability of a transmitter located at $x$, given by
\begin{align} \label{equ:Eqx}
\mathbb{E}[ a_x \gamma_x ] &= \mathcal{Z}\big( \frac{ x^\alpha }{ T r^\alpha } \big) \min\Big\{ \frac{ \xi }{ \mathbb{E}[ \mu^\Phi_0 ] \mathcal{Z}\big( \frac{ x^\alpha }{ T r^\alpha } \big) }, 1\Big\}
\nonumber\\
&= \min\Big\{ \frac{ \xi }{ \mathbb{E}[ \mu^\Phi_0 ] }, \mathcal{Z}\big( \frac{ x^\alpha }{ T r^\alpha } \big) \Big\}.
\end{align}
The result then follows from substituting \eqref{equ:Eqx} into \eqref{equ:Emu0} and manipulating with further algebraic operations.
\end{IEEEproof}

Using the above result, we give a condition for the network to be stable.
\begin{corollary}\textit{In order to maintain the queueing stability of the network, packet arrival rate shall satisfy the following
	\begin{align}
	\xi \leq \mathbb{E}[ \mu^\Phi_0 ] \! \int_0^1 \!\! \mathbb{P}(\eta_{ \mathrm S } > \kappa ) \, d \kappa.
	\end{align}
}
\end{corollary}
\begin{IEEEproof}
By using Loynes' theorem \cite{Loy:62} to the Geo/G/1 queue at the typical link, we have
\begin{align}\label{equ:LoyCndtion}
\xi \leq \mathbb{E}[ \eta_{ \mathrm S } \mu^\Phi_0 ].
\end{align}
Since $\eta_{\mathrm S}$ is constructed under the dominant system, it is thus independent of $\mu^\Phi_0$.
As such, by individually taking expectations to the two random variables on the right hand side of \eqref{equ:LoyCndtion}, the result follows.
\end{IEEEproof}

\subsection{Peak AoI}
We finally obtain the expression for the peak AoI, which characterizes the freshness of information delivery in the wireless network.
\begin{theorem}\textit{If every transmitter makes its transmission decision according to Theorem~\ref{thm:DmSym_PeakAoI}, the peak AoI achievable at the typical link can be computed as
  \begin{align} \label{equ:Calclt_PkAoI}
  A^{\mathrm{p}} \! &= \! \frac{1}{\xi} \!+ \!\! \int_{\xi}^1 \!\!\! \int_{ \xi / v }^1 \frac{ 1 - \xi }{ u v - \xi } F_{\mu}(du) F_{\eta}(dv) \\
  &\approx \frac{1}{\xi} \!+ \! \frac{ 1 - \xi }{ \mathbb{E}[\eta_{ \mathrm{S} }] \mathbb{E}[\mu_{ 0 }^\Phi] - \xi }
  \end{align}
  where $F_{\mu}(\cdot)$ and $\bar{F}_{\eta}(\cdot) = 1 - F_{\eta}(\cdot)$ are given by \eqref{equ:Meta_Grl} and \eqref{equ:SchPlc_Dist}, respectively.
}
\end{theorem}
\begin{IEEEproof}
With the stable condition being satisfied, taking an expectation on both sides of \eqref{equ:CondPeakAoI} yields
\begin{align} \label{equ:Cmpt_PkAoI}
A^{\mathrm{p}} = \frac{1}{\xi} \!+\! \mathbb{E}^0_{\Phi}\! \bigg[ \mathbb{E}\Big[  \frac{ 1 - \xi }{ \gamma^\Phi_0 \mu^\Phi_0 - \xi } \big | \gamma^\Phi_0 \mu^\Phi_0 > \xi \Big] \bigg].
\end{align}
The expression in \eqref{equ:Calclt_PkAoI} then follows from substituting \eqref{equ:SchPlc_Dist} and \eqref{equ:Meta_Grl} in equation \eqref{equ:Cmpt_PkAoI}.
\end{IEEEproof}

Equation \eqref{equ:Calclt_PkAoI} quantifies how all the key features of a wireless network, i.e., interference, scheduling policy, and spatially queueing interaction, affect the AoI.
Several numerical results based on \eqref{equ:Calclt_PkAoI} will be shown in Section V to give more practical insights into the optimization of AoI in a wireless network.

% ================================== %
%              Numerical             %
% ================================== %
\section{Simulation and Numerical Results}
In this section, we verify the accuracy of our analysis through simulations and evaluate the effectiveness of the proposed scheduling policy.
Particularly, during each simulation run, we realize the locations of the transmitters and receivers over a 10 $\text{km}^2$ area via independent PPPs. The packets arriving at each node are generated according to independent Bernoulli processes.
We average over 10,000 realizations and collect the statistic from each communication link to finally calculate the peak AoI.
Unless differently specified, we use the following parameters: $\alpha = 3.8$, $\xi=0.3$, $T=0$~dB, $P_{\mathrm{tx}}=23.7$~dBm, $\sigma^2 = -90$~dBm, and $\lambda = 10^{-4}\mathrm{m}^{-2}$.
\subsection{Validation of Analysis}
\begin{figure}[t!]
  \centering{}

    {\includegraphics[width=0.95\columnwidth]{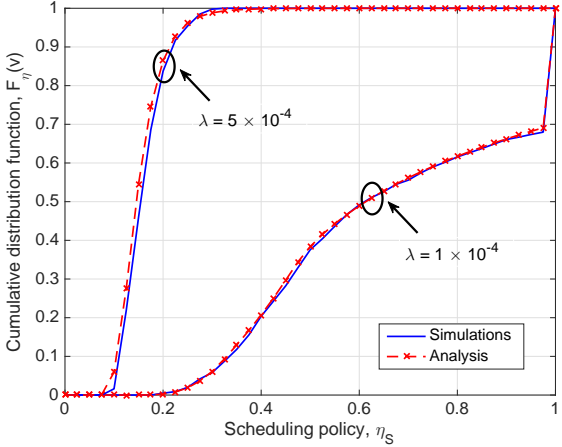}}

  \caption{ CDF of the scheduling policy under deterministic stopping set $S=B(0,R)$, with $R=200$: the transceiver distance is $r=50$, and we vary the spatial density as $\lambda = 1 \times 10^{-4}$ and $\lambda = 5 \times 10^{-4}$. }
  \label{fig:CDF_Schdl}
\end{figure}

Fig.~\ref{fig:CDF_Schdl} depicts the CDF of the scheduling policy under different network densities. We first note the close match between the analysis and simulation results, which validates the analysis in Theorem~\ref{thm:etaS_distribution}.
We further observe that as the spatial density increases, the scheduling policy reduces the channel access probability at different transmitters to keep the mutual interference at a low level.
In this manner, our scheme is able to adjust the access of the radio channel according to the change of the spatial topology.

\begin{figure}[t!]
  \centering{}

    {\includegraphics[width=0.95\columnwidth]{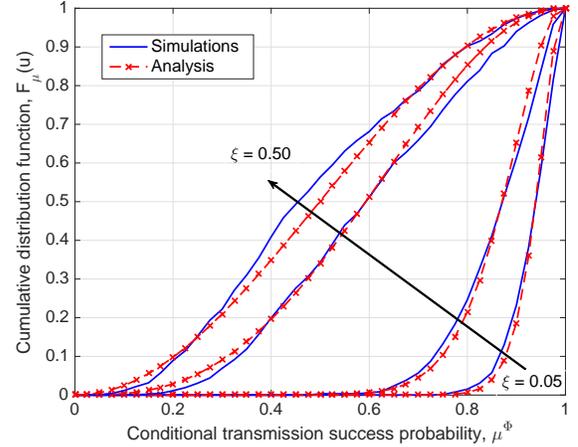}}

  \caption{ CDF of the conditional transmission success probability under deterministic stopping set $S=B(0,R)$, with $R=200$: the transceiver distance is $r=50$, and we vary the packet arrival rate as $\xi=0.05, 0.1, 0.3, 0.5$. }
  \label{fig:CDF_SINR_Schdl}
\end{figure}
In Fig.~\ref{fig:CDF_SINR_Schdl}, we compare the simulated CDF of the conditional transmission success probability to the analysis developed in Theorem~3 for various values of the packet arrival rate $\xi$.
First, the results show a good match for all values of $\xi$, which verify the mathematical analysis.
Next, we can see that an increase in the packet arrival rate defects the conditional transmission success probability, or equivalently, service rate, in a non-linear manner.
Specifically, the service rate declines rapidly as the traffic condition changes from low ($\xi=0.05$) to medium ($\xi=0.3$) regime, while the trend slows down as the packet arrival rate further goes up ($\xi=0.5$).
This mainly amounts to the composite effect of the temporal variation: in the light traffic condition, when the packet arrival rate goes up, the accrued packets at each buffer will incur additional transmission attempts. In consequence, more transmitters are activated and they together raise the interference level.
As a result, the received SINR at each node decreases and the active duration of transmitters is then prolonged, which in turn defects the service rate across the network.
In the heavy traffic regime, however, as the majority of the queues are already saturated, the additional activated transmitters cannot largely change the interference, and thus the descent of conditional transmission success probability is leveled off.

\subsection{ Performance Evaluation }
\begin{figure}[t!]
  \centering{}

    {\includegraphics[width=0.95\columnwidth]{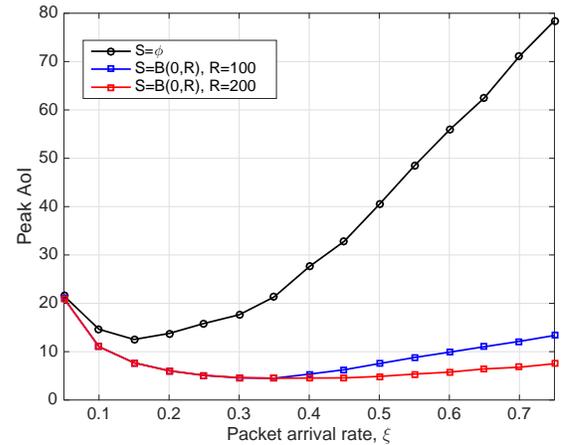}}

  \caption{ Peak AoI vs packet arrival rate: $r=25$, with deterministic stopping set $S=B(0,R)$, where $R=100$. }
  \label{fig:PAoI_SmlR}
\end{figure}
We now compare the proposed scheduling scheme with local observation from a deterministic stopping set to that with no available local information, i.e., $S=\phi$ (in which case, $\eta_{\mathrm{S}} =1$, $\forall j \in \mathbb{N}$) in Fig.~\ref{fig:PAoI_SmlR}.
From this figure, we immediately note an optimal packet arrival rate exists for both cases due to a tradeoff between update frequency and the incurred delay.
Moreover, the figure also shows that once armed with sufficient local information, the proposed scheduling method is able to maintain the AoI at a low level for a wide range of packet arrival rates, demonstrating its effectiveness in optimizing the information freshness in wireless networks.

\begin{figure}[t!]
  \centering{}

   \subfigure[\label{fig:1a_prim}]{\includegraphics[width=0.95\columnwidth]{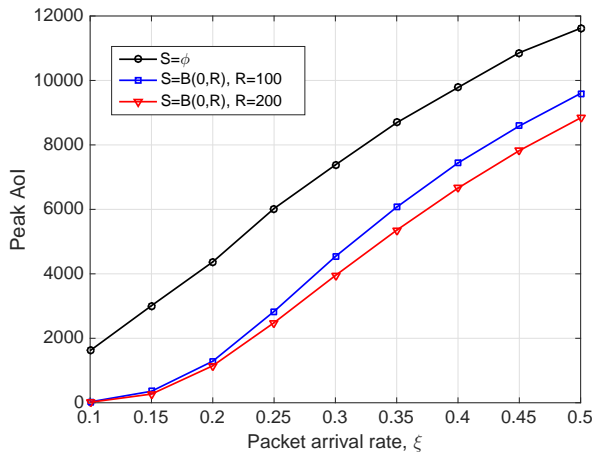}}
   \subfigure[\label{fig:1b_prim}]{\includegraphics[width=0.95\columnwidth]{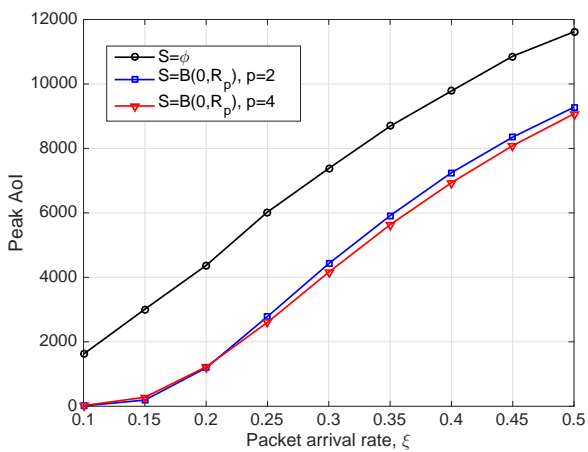}}

  \caption{ Peak AoI vs packet arrival rate: $r=100$, (a) peak AoI as a function of $\xi$, for deterministic stopping set $S = B(0,R)$, (b) peak AoI as a function of $\xi$, for random stopping set $S=B(0,R_p)$. }
  \label{fig:PeakAoI_vs_xi}
\end{figure}
Fig.~\ref{fig:PeakAoI_vs_xi} further shows the peak AoI per packet arrival rate under scenarios with no available local information, i.e., $S=\phi$, and that with different forms of stopping sets.
In particular, we consider the stopping set taking both deterministic form, i.e., $S=B(0,R)$ with $R$ being a constant, and random form, i.e., $S=B(0,R_p)$, where $R_p$ denotes the distance to the $p$-th closest receiver from a typical transmitter.
The figure carries multiple consequential messages:
\begin{itemize}
\item In spite of suffering an SINR degradation, the proposed scheduling policy greatly improves the peak AoI compared to that without local information. The gain is especially remarkable in the regime with low to moderate packet arrival rates, in which the proposed scheme reduces more than half of the peak AoI. This is mainly because this is the regime where the SINR rapidly degrades while only a few packets are accumulated in the buffer, and hence if transmitters can tolerate certain increments in the queueing delay and control their channel access frequency, the SINR can be greatly boosted up at each node which leads to a much shorter transmission delay.

\item As long as the similar amount of local information can be extracted to make the transmission decision, there is little difference in the performance gain attained from using deterministic or random stopping sets. This observation implies that it is the amount of information that matters to the design of a scheduling policy rather than the specific shape of the observation window, which is in line with Remark~3.

\item In both instances, the performance of the proposed scheme is shown to be enhanced through expanding the observation region, i.e., by increasing the radius of the deterministic disk (Fig.~\ref{fig:1a_prim} in this case) or directly counting in more geographical neighbors (Fig.~\ref{fig:1b_prim} in this case).
    Theoretically, such expansion can keep increasing indefinitely to make the performance of our local scheduling policy ultimately reach that of a central controller, while practically, the gain from extending the observation window is marginal compared to the additional complexity and is thus not desirable.

\item Nevertheless, note that even availing the transmitters with information from one or two neighbors, it is able to tremendously reduce the peak AoI by using our proposed scheduling scheme.
This observation demonstrates the practical effectiveness of our approach.
\end{itemize}

\begin{figure}[t!]
  \centering{}

    {\includegraphics[width=0.95\columnwidth]{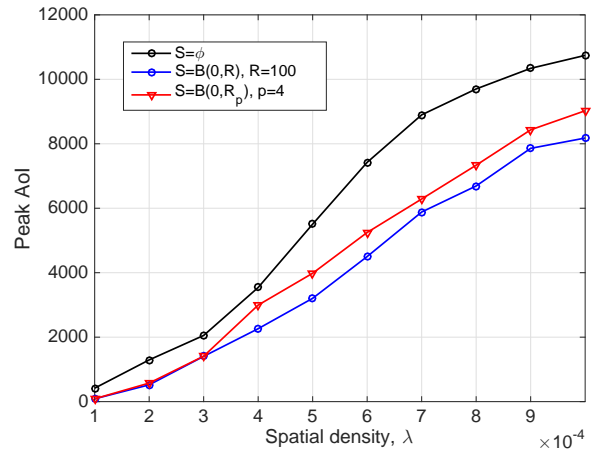}}

  \caption{ Peak AoI vs spatial density: $r=25$, $\xi=0.3$, $p=4$ for the random stopping set $S=B(0,R_p)$, and $R=100$ for the deterministic stopping set $S=B(0,R)$. }
  \label{fig:PeakAoI_vs_lambda}
\end{figure}

Fig.~\ref{fig:PeakAoI_vs_lambda} depicts the peak AoI as a function of the spatial density for scenarios with and without local observation in the scheduling policy design.
This figure not only illustrates how network densification affects the information freshness, but also highlights the critical role played by the scheduling policy. We hence conclude the observation into the following takeaways:
\begin{itemize}
\item The peak AoI always increases with respect to the spatial density, since densifying the network inevitably entails additional interference, thus the SINR is defected and it further hurts the transmission quality across network.

\item By employing the locally adaptive scheduling policy at each transmitter, the peak AoI undergoes a substantial discount and the gain is more pronounced in the dense network scenario.
This is because the interference between neighbors becomes more severe when their mutual distance is reduced, and hence adequately scheduling the channel access patterns of transmitters can prevent interference from rising too quickly and maintain the peak AoI at a low level.

\item When adopting a deterministic stopping set, the proposed scheduling policy benefits more from network densification. The reason comes from the fact that by fixing the size of the observation window, information from more neighbors can be taken into account when the spatial density increases, thus allowing a transmitter to assert better response. Such observation is also inline with the above discussion.
\end{itemize}

% ================================== %
%            Conclusion              %
% ================================== %
\section{Conclusion}
In this paper, we conducted an analytical study on the design of a scheduling policy that optimizes information freshness in wireless networks.
We proposed a decentralized protocol that allows every transmitter to make transmission decisions based on the observed local information.
Using the concept of stopping sets, we encapsulated the local knowledge from individual nodes in the analytical framework, and derived tractable expressions to characterize the stochastic behavior of our proposed scheme, as well as quantified its effectiveness in terms of peak AoI.
Numerical results showed that while the link throughput is generally affected by the packet arrival rate, our proposed scheme managed to adapt the transmission to the traffic variation and hence largely reduced the peak AoI.
Moreover, the scheme has also been shown to adaptively adjust according to the geographical change of the ambient environment and thus scales well as the network grows in size.

By combining queueing theory with stochastic geometry, the developed framework bridges the gap between the abstract service model -- which is widely used in the existing AoI literature -- and the physical transmission environment.
It thus enables one to devise fundamental insights on the impact from both spatial and temporal aspects of a network on the information freshness. The model can be further applied in the design of scheduling schemes in wireless systems under different queueing disciplines, or with multiple sources as well as multi-hop routings. Investigating up to what extent a non-binary power control can improve AoI is also regarded as a concrete direction for future work.

\begin{appendix}
\section{Proof of Lemma~\ref{lma:CndtSucProb_Dmnt}} \label{apx:CndtSucProb_Dmnt}
Because every link is backlogged in the dominant system, the SINR received at the typical link can be written as
\begin{align}
\mathrm{SINR}^{\mathrm{D}}_{0} = \frac{ P_{ \mathrm{tx} } H_{00} r^{-\alpha} }{ \sum_{ j \neq 0 } { P_{ \mathrm{tx} } H_{j0} \nu_{j} }{ \Vert X_j \Vert^{-\alpha} } \!+\! \sigma^2 }.
\end{align}
Therefore, by conditioning on the spatial realization $\Phi$ of all the links, we can compute the transmission success probability as follows:
\begin{align}
\hat{\mu}^\Phi_0 &=\mathbb{P}\!\left( \mathrm{SINR}^{\mathrm D}_{0} \!>\! T | \Phi  \right)
\nonumber\\
&= \mathbb{P} \bigg(  { H_{00} }  > { T r^\alpha } \Big( \sum_{ j \neq 0 } \frac{ H_{j0} \nu_{j} }{ \Vert X_j \Vert^\alpha } \!+\! \frac{1}{\rho} \Big) \,  \Big| \, \Phi  \bigg)
\nonumber\\
&=  \mathbb{E}\bigg[ e^{-\frac{ T r^\alpha }{\rho}} \prod_{ j \neq 0 } \exp\! \Big(\! - T r^\alpha \frac{ H_{j0} \nu_{j} }{ \Vert X_j \Vert^\alpha } \Big)  \Big| \Phi \bigg]
\nonumber\\
&\stackrel{(a)}{=}   e^{-\frac{ T r^\alpha }{\rho}} \prod_{ j \neq 0 }  \Big( 1 - \gamma_j^\Phi - \gamma_j^\Phi \mathbb{E}\big[ \exp \big(\frac{ H_{j0} }{ \mathcal{D}_{j0} } \big) \big]  \Big),
\end{align}
where $(a)$ follows by noticing that the random variables $\{ H_{j0}, j = 1, 2, ... \}$ are i.i.d. and the scheduling policy is constructed on $\Phi$, namely $\gamma_j^\Phi = \eta_S( \theta_{X_j} \tilde{\Phi}, \theta_{X_j} \bar{\Phi} )$.
The result then follows from further algebraic manipulations.

\section{Proof of Theorem~\ref{thm:DmSym_PeakAoI}} \label{apx:OptCtrl_PeakAoI}
First of all, we note that under a dominant system, the point process $\Phi$ is stationary.
As such, by substituting \eqref{equ:Dmnt_SucProb} into the first term of \eqref{prbm:Dom_PeakAoI} and using the mass transportation theorem \cite{BacBla:09}, we obtain the following:
\begin{align}\label{equ:MasTrn}
&\mathbb{E}^0_\Phi \!\! \left[ \frac{ 1 - \xi }{  \gamma^\Phi_0 \mu^\Phi_0 - \xi  } \right] \!=\!  \mathbb{E}^0_\Phi \! \bigg[ \frac{ 1 - \xi }{ \eta_{\mathrm{S}} \! \prod_{j \neq 0} \! \left( 1 \! - \! \frac{ \eta_{\mathrm{S}} }{ 1 + \mathcal{D}_{0j} } \right) \! e^{\frac{ - T r^\alpha }{\rho}} \!\!\!-\! \xi  } \bigg].
\end{align}
Our goal is now to minimize the above expression as a function of $\eta_{\mathrm{S}}$ under the constraint in \eqref{equ:Dom_Rst_SpSt}. To accomplish this target, we separate the denominator into two sets depending on whether a generic receiver $y_i \in S$ or not. As such, by the strong Markov property, \eqref{equ:MasTrn} can be written as follows
\begin{align}\label{equ:SpltDnm}
&\mathbb{E}^0_\Phi \! \bigg[ \frac{ 1 - \xi }{ \eta_{\mathrm{S}}  \prod_{j \neq 0} \! \left( 1 \! - \! \frac{ \eta_{\mathrm{S}} }{ 1 + \mathcal{D}_{0j} } \right) \! e^{-\frac{ T r^\alpha }{\rho}} \!\!\!-\! \xi  } \bigg]
\nonumber\\
= \, & \mathbb{E}^0_\Phi \! \bigg[ \frac{ (1-\xi) e^{ T r^\alpha / \rho } }{ \eta_{\mathrm{S}} \!\!\!\!\!\! \prod\limits_{ \substack{ j \neq 0,  y_j \in S } } \!\!\!\!\!\! ( \, 1 \! - \! \frac{ \eta_{\mathrm{S}} }{ 1 + \mathcal{D}_{0j} } \, ) \, e^{ -\! \int_{ \mathbb{R}^2 \! \setminus \! S } \!\!\! \frac{ \lambda \eta_{\mathrm{S}}  dy }{ 1 + y^\alpha \!/ T r^\alpha } } \!+ \xi e^{ T r^\alpha / \rho } } \bigg].
\end{align}
The minimization of \eqref{prbm:Dom_PeakAoI} now becomes minimizing the above expression with respect to $\eta_{\mathrm{S}}$.
Note that as $\eta_{\mathrm{S}}$ is defined as a function, \eqref{prbm:Dom_PeakAoI} is in essence a functional optimization problem which should be solved by the calculus of variations. Fortunately, as the operator $\eta_{\mathrm{S}}$ is well-defined in the stationary point process $\Phi$, with the help of \eqref{equ:SpltDnm}, we can treate the operator as a variable \cite{BacBlaSin:14,RamBac:19} and take the derivative of \eqref{prbm:Dom_PeakAoI} with respect to $\eta_{\mathrm{S}}$ and equate it to zero, which yields the following:
\begin{align} \label{equ:FixPnt_PRF}
\frac{1}{\eta_{\mathrm{S}}} -\!\!\!\!\!\! \sum_{ \substack{ j \neq 0, y_j \in S } }  \frac{1}{ 1 \!+\! \mathcal{D}_{0j}   \!-\! \eta_{\mathrm{S}} } -\!\! \int_{\mathbb{R}^2 \setminus S }\! \frac{ \lambda \,  dy }{ 1 \!+\! y^\alpha \! / T r^\alpha } = 0.
\end{align}
If we write the left hand side (L.H.S.) of the above equation as a function $f(\eta_S)$ of $\eta_S$, i.e.,
\begin{align}
f(\eta_S) = \frac{1}{\eta_{\mathrm{S}}} -\!\!\!\!\! \sum_{ \substack{ j \neq 0, y_j \in S } } \frac{1}{ 1  \!+\! \mathcal{D}_{0j} \!-\! \eta_{\mathrm{S}} } -\!\! \int_{\mathbb{R}^2 \setminus S }\! \frac{ \lambda  dz }{ 1 \!+\! \Vert z \Vert^\alpha \! / T r^\alpha },
\end{align}
it is easy to verify that ($a$) $f(\eta_S)$ monotonically decreases in $\eta_S$ over [0, 1], and ($b$) $\lim_{ \eta_S \rightarrow 0 } f(\eta_S) = +\infty$. As such, if $f(1) < 0$, i.e., the condition \eqref{equ:CndOptl} holds, then according to the Intermediate Value Theorem, the equation in \eqref{equ:FixPnt_PRF}, or equivalently, (15), has a solution and this solution is unique.
 % is continuous and decreasing in $\eta_{\mathrm{S}}$ over [0, 1], thus the equation has one solution if (16) holds.
Otherwise, if \eqref{equ:CndOptl} does not hold, we have the derivative of \eqref{equ:SpltDnm} being negative which indicates that \eqref{prbm:Dom_PeakAoI} monotonically decreases as a function of $\eta_{\mathrm{S}}$. Hence, the minimum is achieved at $\eta_{\mathrm{S}} = 1$.

\section{Proof of Theorem~\ref{thm:etaS_distribution}} \label{apx:etaS_distribution}
From the argument in \eqref{equ:FixPnt_PRF}, we know that for all $0 < \kappa < 1$, there is $\eta_{\mathrm{S}} > \kappa$ as long as the following relationship holds
\begin{align*}
\sum_{ j \neq 0, y_j \in S }  \frac{\kappa}{ 1 \!+\! \mathcal{D}_{0j}  \!-\! \kappa } +\!\! \int_{ \mathbb{R}^2 \setminus S }\! \frac{ \lambda \kappa dy }{ 1 \!+\! \Vert y \Vert^\alpha \! /  T r^\alpha } < 1,
\end{align*}
which can be equivalently written as $\mathcal{U}(\kappa, S) + \mathcal{V}(\kappa, S) < 1$, where $\mathcal{U}( \kappa, S )$ and $\mathcal{V}( \kappa, S )$ are given by \eqref{equ:UkS} and \eqref{equ:VkS}, respectively.
As such, by using Slivnyark's theorem \cite{BacBla:09}, we have
\begin{align*}
\mathbb{P}\left( \eta_{\mathrm{S}} > \kappa | \Phi \right) &= \mathbb{P}^0\left(\, \mathcal{U}(\kappa, S) < 1 - \mathcal{V}(\kappa, S) \,\right)
\nonumber\\
&= \mathbb{P}\left(\, \mathcal{U}(\kappa, S) < 1 - \mathcal{V}(\kappa, S) \, \right).
\end{align*}
Using this result, we can thus write the CCDF of $\eta_{\mathrm{S}}$ as
\begin{align*}
\mathbb{P}\! \left( \eta_{\mathrm{S}} > \kappa \right) = \! \int_{0}^{ 1 - \mathcal{V}(\kappa, S) } \!\!\!\!\!\!\!\!\!\!\!\!\!\!\!\!  g_{\kappa}(u) du = \! \int_{-\infty}^{+\infty} \!\!\!\!\!\! g_{\kappa}(u) h_{\kappa}(u) du
\end{align*}
where $g_{\kappa}(u)$ is the probability density function (PDF) of the random variable $\mathcal{U}{ (\kappa,S) }$, and $h_{\kappa}(u)$ is an indicator function that takes value 1 if $0 \leq u \leq 1 - \mathcal{V}(\kappa, S)$, and 0 otherwise. As such, by applying the Plancherel-Parseval theorem \cite{BacBla:09} to the above equation, we arrive at the following expression
\begin{align} \label{equ:Prob_etaS}
\mathbb{P}\! \left( \eta_{\mathrm{S}} > \kappa \right) = \frac{ 1 }{ 2 \pi } \int_{-\infty}^{+\infty} \mathcal{F}_{\mathcal{U}(\kappa,S)}(\omega) \mathcal{F}^*_{ \mathcal{V}(\kappa,S) }(\omega) d \omega
\end{align}
where $\mathcal{F}_A(\omega) = \mathbb{E}[ \exp( - j \omega A ) ]$ is the characteristic function of a random variable $A$ and $\mathcal{F}^*_A(\omega)$ denotes the corresponding complex conjugate.
To this end, we first attain the following result for $\mathcal{V}(\kappa, S)$, i.e.,
\begin{align} \label{equ:F_conjA}
\mathcal{F}^*_{ \mathcal{V}(\kappa,S) }(\omega) = { \left[ e^{ j \omega ( 1 - \mathcal{V}(\kappa, S) ) } - 1 \right] }/{ j \omega }.
\end{align}
On the other hand, the characteristic function $\mathcal{F}_{\mathcal{U}(\kappa,S)}(\omega)$ can be derived from its Laplace transform, which is given as
\begin{align} \label{equ:LT_UkS}
&\mathcal{L}_{ \mathcal{U}(\kappa,S) }(s) \!=\! \mathbb{E}\!\left[ \prod_{ y_j \in S }\!\!  \exp\Big({ - \frac{ s \kappa }{ 1 \!-\! \kappa + \mathcal{D}_{0j}  } } \Big)  \! \right]\!.
\end{align}
The theorem then follows from taking \eqref{equ:F_conjA} and \eqref{equ:LT_UkS} into \eqref{equ:Prob_etaS} and performing further algebraic manipulation.

\section{Proof of Lemma~\ref{lma:CndSlct} }\label{apx:CndSlct}
Using a similar approach as in the proof of Theorem~\ref{thm:etaS_distribution}, we have that for all $0 < \kappa < 1$, in order to achieve $\eta_{ \mathrm S }( \theta_{ X_j } \tilde{\Phi}, \theta_{ X_j } \bar{\Phi} ) > \kappa$, we will need the following condition to hold:
\begin{align} \label{equ:CndCnd}
\sum_{ i \neq j, y_i \in S } \frac{ \kappa }{ 1 + \mathcal{D}_{ ji } - \kappa } + \int_{ \mathbb{R}^2 \setminus S } \frac{ \lambda \kappa dy }{ 1 + \Vert y \Vert^\alpha /T r^\alpha } < 1.
\end{align}
Now, given a receiver located at the origin, we can rewrite \eqref{equ:CndCnd} in the following way:
\begin{align}
\sum_{ \substack{ i \neq j, i \neq 0, \\ y_i \in S} } \!\! \frac{ \kappa }{ 1 \!+\! \mathcal{D}_{ji} \!-\! \kappa } \!<\! 1 \!-\! \mathcal{V}( \kappa, S ) - \frac{ \kappa }{ 1 \!+\! \mathcal{D}_{j0} \!-\! \kappa }.
\end{align}
As such, the conditional distribution of the scheduling policy can be attained by a similar manner as in the proof of Theorem~\ref{thm:etaS_distribution}, given by
\begin{align} \label{equ:Cnd_CndSlctProb}
&\mathbb{P}( \eta_{ \mathrm S } > \kappa \big| \zeta_j = 1,  \Vert X_j - y_0 \Vert = l )
\nonumber\\
&\stackrel{(a)}{=} \! \int_{-\infty}^{+\infty} \!\!\!\!\!\!\!\! \mathcal{L}_{\mathcal{U}(\kappa, S)} (j \omega) \, \frac{ e^{j \omega ( 1 - \mathcal{V}( \kappa, S ) - \frac{ \kappa }{ 1 - \kappa + l^\alpha \!/ T r^\alpha } ) } - 1 }{ j 2 \pi \omega } d \omega.
\end{align}
where ($a$) follows from Slivnyark's theorem. Finally, using the above, we have that
\begin{align} \label{equ:ProofCnd_SclActProb}
& \mathbb{P}( \nu_j = 1 | \zeta_j = 1, \Vert X_j - y_0 \Vert = l )
\nonumber\\
&\qquad \qquad  = \int_0^1 \mathbb{P}( \eta_{ \mathrm S } > \kappa \big| \zeta_j = 1,  \Vert X_j - y_0 \Vert = l ) d \kappa.
\end{align}
Lemma~\ref{lma:CndSlct} then immediately follows by substituting \eqref{equ:Cnd_CndSlctProb} into \eqref{equ:ProofCnd_SclActProb}.

\section{Proof of Theorem~\ref{thm:SINR}} \label{apx:SINR_proof}
Under Assumption~1, we can focus on the steady state of the network and drop the time index in the subsequential analysis.
To facilitate the presentation, we introduce two notations $Y^\Phi_i$ and $q_i$, defined as $Y^\Phi_i = \ln \mathbb{P}(\mathrm{SINR}_i > T | \Phi)$ and $q_{l,i} = \mathbb{P}( \nu_{i} \times  \zeta_{i} = 1 | \Vert X_i - y_0 \Vert = l )$, respectively.
Using Slivnyark's theorem \cite{BacBla:09}, we concentrate on the moment generating function of $Y^\Phi_0$ as follows:
\begin{align} \label{equ:M_Y0}
&M_{Y^\Phi_0}(s) = \mathbb{E}\left[ \exp\left( s Y^\Phi_0 \right) \right] = \mathbb{E}\left[ \mathbb{P}\left( \mathrm{SINR}_0 > T | \Phi \right)^s \right]
\nonumber\\
&\stackrel{(a)}{=} \mathbb{E}\!\left[ e^{ - \frac{ s T r^\alpha }{ \rho } } \prod_{ i \neq 0 } \! \left( 1 \!-\! q_i \!+\! \frac{ q_i }{ 1 + T r^\alpha \Vert X_i \Vert^\alpha } \right)^s \right]
\nonumber\\
&\stackrel{(b)}{=} \exp\left( - 2\pi \lambda \int_0^{\infty} \sum_{k=1}^s (-1)^{k+1} \frac{ \mathbb{E}[q_x^k] x dx }{ 1 + x^\alpha / T r^\alpha }  \right),
\end{align}
where ($a$) follows from the independent evolution of queues according to Assumption~1, and ($b$) by using the probability generating functional (PGFL).

The complete expression of \eqref{equ:M_Y0} requires us to further compute $\mathbb{E}[q_{x}^k]$, which can be written as
\begin{align} \label{equ:q_x_k}
\mathbb{E}[q_x^k] & \! \stackrel{(a)}{=} \! \mathbb{E}\Big[ \Big( \gamma^\Phi_x \cdot \min\Big\{ \frac{ \xi }{ \mu^\Phi_x \gamma^\Phi_x }, 1 \Big\} \Big)^k \Big]
\nonumber\\
& = \mathbb{E}\Big[ \! \Big( \! \left( \gamma^\Phi_x \right)^k \chi_{ \{ \gamma^\Phi_x \mu^\Phi_x < \xi \} } \!+\! \Big( \frac{ \xi }{ \mu^\Phi_x } \Big)^k \! \chi_{ \{ \gamma^\Phi_x \mu^\Phi_x \geq \xi \} } \Big)  \Big],
\end{align}
where ($a$) follows from the Little's Law \cite{Har:13}.

At this stage, let us temporally assume the CDF $F_\mu(u)$ of $\mu^\Phi_0$ is available. Using Lemma~4, the first term of \eqref{equ:q_x_k} can then be derived as
\begin{align} \label{equ:qx_term1}
\mathbb{E}\Big[ \! \left( \gamma^\Phi_x \right)^k \chi_{ \{ \gamma^\Phi_x \mu^\Phi_x < \xi \} }   \Big] = \mathcal{Z}^k(x) F_\mu( \xi / \mathcal{Z}(x) ).
\end{align}
Similarly, we have the second term of \eqref{equ:q_x_k} given by
\begin{align} \label{equ:qx_term2}
\mathbb{E}\Big[ \!\Big( \frac{ \xi }{ \mu^\Phi_x } \Big)^k  \chi_{ \{ \gamma^\Phi_x \mu^\Phi_x \geq \xi \} }   \Big] = \int_{ \xi / \mathcal{Z}(x) } \frac{\xi^k}{ t^k } F_{\mu}(dt).
\end{align}

Finally, we apply the Gil-Pelaez theorem \cite{Gil} and derive the CDF of $\mu^\Phi_0$ as
\begin{align} \label{equ:Gil_Pel}
F_{\mu}(u) &= \mathbb{P}\left( \mu^\Phi_0 < u \right) = \mathbb{P}\left( Y_{0}^\Phi < \ln u \right)
\nonumber\\
&= \frac{1}{2} - \frac{1}{\pi} \! \int_0^\infty \! \mathrm{Im}\left\{ u^{- j \omega}  M_{ Y_{0}^\Phi }(j \omega) \right\} \frac{d \omega}{\omega}.
\end{align}
To this end, by substituting \eqref{equ:q_x_k}, \eqref{equ:qx_term1}, and \eqref{equ:qx_term2} into the above equation, the result follows.
\end{appendix}

\bibliographystyle{IEEEtran}
%% argument is your BibTeX string definitions and bibliography database(s)
\bibliography{bib/StringDefinitions,bib/IEEEabrv,bib/howard_AoI_Schdl}

\end{document}